\documentclass[preprint,10pt]{iopart}
\pdfoutput=1
\usepackage{iopams} 
\usepackage{graphicx}
\usepackage{color}

\begin{document}
\textbf{{\color{red}Accepted for publications in J. Phys.: Condens. Matter (2012)}}

\title[]{Ab initio prediction of pressure-induced structural phase transition of superconducting FeSe}

\author{Gul Rahman }
\address{Department of Physics,
Quaid-i-Azam University, Islamabad 45320, Pakistan}

\author{In Gee Kim}
\address{Graduate Institute of
Ferrous Technology, Pohang University of Science and Technology, Pohang 790-784, Republic of Korea}
\ead{igkim@postech.ac.kr}

\author{Arthur J. Freeman}
\address{Department of Physics and Astronomy,
Northwestern University, Evanston, IL 60208, U. S. A.}

\begin{abstract}
External pressure driven phase transitions of FeSe are predicted using \textit{ab initio} calculations. The calculations reveal that $\alpha$-FeSe takes transitions to NiAs-type, MnP-type, and CsCl-type FeSe. Transitions from NiAs-type to MnP-type and CsCl-type FeSe is also predicted. MnP-type FeSe is also found to be able to transform to CsCl-type FeSe, which is easier from $\alpha$-FeSe than the transition to MnP-type FeSe, but comparable to the transition from NiAs-type FeSe. The calculated electronic structures show that all  phases of FeSe are metallic, but the ionic interaction between Fe-Se bonds becomes stronger and the covalent interaction becomes weaker when the structural phase transition occurs from $\alpha$-FeSe to the other phases of FeSe. The experimentally observed decrease in $T_{c}$ of superconducting $\alpha$-FeSe at high pressure may be due to a structural/magnetic instability, which exists at high pressure. The results suggest us to increase the $T_{c}$ of $\alpha$-FeSe if such phase transitions are frustrated by suitable methods.
\end{abstract}

\pacs{71.20.-b, 74.62.Fj, 74.70.-b}

\maketitle

\section{Introduction}
\label{sec:intro}

FeSe has been studied for its magnetic properties~\cite{Wu,Wu2,gulJKPS} and showed a phase transition from $\alpha$-FeSe to $\beta$-FeSe (tetragonal to hexagonal) as the growth temperatures were increased~\cite{Wu2}. Interest in $\alpha$-FeSe have grown after Hu \textit{et al.}~\cite{Hu} observed superconductivity with a $T_c \sim 8$\;{K} because it is structurally related to the newly existing investigations of FeAs superconductors~\cite{JACS} and subsequently the $T_{c}$ was increased by the application of pressure~\cite{Medv,Mizu} and doping~\cite{FeTe-ref}. External pressure up to $2.2$\;{GPa} enhances both spin fluctuations and $T_{c}$ in $\alpha$-FeSe~\cite{Imai}; however a large external pressure further decreases $T_{c}$~\cite{Medv}.

Experimental results show that tetragonal FeSe  undergoes a structural transition to a hexagonal phase~\cite{Brai} and at $12.5$\;{GPa} to an orthorhombic phase~\cite{Garb}.  More recently, Medvedev \textit{et al.}~\cite{Medv} reported the increase of $T_{c}$ from $8.5$ to $36.7$\;{K} under an applied pressure of $8.9$\;{GPa} and also showed a phase transition from tetragonal to NiAs-type hexagonal at $12$\;{GPa}. Compared with the FeAs superconductors~\cite{JACS}, which also show a structural phase transition from tetragonal to orthorhombic~\cite{Zhao,Namura}, $\alpha$-FeSe has not only  the same planar sublattices, but also displays structural and magnetic instabilities~\cite{Yeh1}.

Albeit FeSe shows structural transitions, it is not clear how the electronic structure responds to external pressure, and what the transition path of FeSe should be. It is  then invaluable to study FeSe in different crystallographic environments to observe the effects of structural changes upon changing the atomic volume or applying pressure.
Thus, because of its comparative chemical simplicity with FeAs superconductors, FeSe provides a unique opportunity to study the interplay of the structure, magnetism, and superconductivity. Therefore, the determination of the structural and magnetic phase transitions in FeSe is essential for understanding their electronic properties. To our knowledge, no systematic study  has been performed which can explore the mechanism of the structural
phase transitions in stoichiometric FeSe. We show that there are structural and magnetic phase transitions at high pressure. Such phase transitions also affect the magnetic, electronic, and bonding properties of FeSe; these are considered as an indirect clue for improving the $T_{c}$ of FeSe.

This paper is organized as follows. We describe the computational models and methods in Sec.~\ref{sec:model} which are used to investigate the structural phase transitions of FeSe. In Sec.~\ref{sec:results}, we discuss our calculated results, \textit{i.e.}, structural phase transition, magnetism, and electronic structures of FeSe. The results and discussions are summarized in Sec.~\ref{sec:conclusion}.

\section{Computation Models and Methods}
\label{sec:model}

To search for the structural phase transition of FeSe, we considered different types of crystal structures, \textit{i.e.}, anti-PbO---commonly known as $\alpha$-FeSe, as well as NiAs-, MnP-, CsCl-, and CuAu-type FeSe (see Fig.~\ref{structures}). These structures are related with each other, \textit{e.g.}, MnP-type FeSe can be considered as a deformed NiAs-type FeSe, changing the local hexagonal symmetry around the Fe atoms in the NiAs-type phase to tetrahedral results in the $\alpha$-FeSe phase. The detailed structure relationship will be discussed  in Sec.~\ref{sec:results}.

\textit{Ab initio} calculations were performed using the total-energy all-electron full-potential linearized augmented plane-wave (FLAPW) method ~\cite{FLAPW} in the QMD-FLAPW package~\cite{FLAPW-web} based on both the local spin density approximation (LSDA)~\cite{LSDA} and the generalized gradient approximation (GGA)~\cite{GGA}. Integrations inside the Brillouin zone (BZ) were performed by the improved tetrahedron method~\cite{TETRA} over a $15\times 15\times 15$ mesh within the three dimensional (3D) BZ and with an energy cutoff at $4.10$\,{$(2\pi/a)$}, where $a$ is the lattice parameter of each calculation. 
A $16.36$\,{$(2\pi/a)$} star-function cutoff was used for depicting the charge and potential in the interstitial regions. 
Lattice harmonics with $l \leq 8$ were employed to expand the charge density, potential, and wave-functions inside each muffin-tin (MT) sphere of radii $2.2$\;{a.u.} for Fe and $1.9$\;{a.u.} for Se. The convergence of the computation parameters were checked carefully~\cite{Seo2009}.

All core electrons were treated fully relativistically and valence states were calculated scalar relativistically, \textit{i.e.}, without spin-orbit coupling~\cite{Rela}. The explicit orthogonalization (XO) procedure was employed to ensure the orthogonality between the semicore and valence states with the spill-out of each MT sphere semicore charges being superposed~\cite{XO}. Self-consistency was assumed when
the difference between input and output charge densities became less than $1.0\times10^{-4}$\,{electrons/a.u.$^{3}$}

The calculated total energy-volume ($E$-$V$) data were fitted to the Birch-Murnaghan equation of states~\cite{birch} which enables us to calculate the enthalpy $H = E + pV$, where $E$ is the internal energy, $p$ is the external pressure, and $V$ is the volume of the system. All calculations were carried out for nonmagnetic (NM) and ferromagnetic (FM) states at different volumes. The optimized lattice parameters of $\alpha$-FeSe were calculated to be $a= 3.747\,(3.701)$\;{\AA}, $c/a= 1.474\,(1.505)$, and volume/atom ratio $=19.685\,(19.078)$\;{\AA$^3$} with the GGA (LSDA) results which are comparable to the recent experimental values~\cite{Garb}.
The calculated internal coordinates and the formula unit volumes 
of the considered crystals at individual equilibria are summarized in Table \ref{Atomic}.
We carefully checked that the physics of these systems does not depend on the details of internal parameter values.

\section{Results and Discussions}
\label{sec:results}
\subsection{Structural Phase Transition}
In Fig.~\ref{T-path}, one can see that $\alpha$-FeSe is the most stable structure and under sufficient volume contraction (external pressure), $\alpha$-FeSe becomes unfavorable compared to the other phases of FeSe. Around the equilibrium volume of $\alpha$-FeSe all other phases are in metastable states by an energy barrier per formula unit (f.u.) of order $1.0$\;{eV}. As we compress the unit cell volume, $\alpha$-FeSe goes to a higher energy state and NiAs-type FeSe becomes a stablized phase. NM $\alpha$-FeSe starts to cross FM NiAs-type FeSe at a volume of $\sim 30.74$\;{\AA$^3$}. Further compression transits $\alpha$-FeSe to the NM NiAs-type phase at $30.19$\;{\AA$^3$}, but before crossing the NM NiAs-type FeSe phase, $\alpha$-FeSe crosses the energy curve of MnP-type FeSe. Hence, a slight compression deforms NiAs-type FeSe to MnP-type FeSe, where the hexagonal symmetry is broken and can be viewed as a distorted NiAs structure. We can also see that NM $\alpha$-FeSe begins to overcome FM MnP-type FeSe at about $30.33$\;{\AA$^3$} and  NM MnP-type FeSe at about $29.86$\;{\AA$^3$}.

NiAs-type FeSe can be transformed to CsCl-type FeSe if we push the Se atom to the center of the unit cell and simultaneously deform the $a$ and $b$ axes of the hexagonal lattice to a cubic lattice under a large external pressure, to break the bonds between Fe and Se atoms in the hexagonal structure for making new bonds in the cubic structure. In other words, tetragonal $\alpha$-FeSe can be thought of as a distorted CsCl-type FeSe with an elongated $c$ axis. Hence, one can naturally expect a transition to CsCl-type FeSe at high pressure corresponding. On the other hand, MnP-type FeSe is a distorted structure where phase transition to CsCl-type FeSe will require larger pressure as compare to NiAs-type FeSe. Although CuAu-type FeSe lies much higher in energy, extrapolating $\alpha$-FeSe to a small volume region tells us that $\alpha$-FeSe can enter into a metastable region at $27.80$\;{\AA$^3$} and CuAu-type FeSe may become a stable structure. However, the common tangent does not pass through both curves which makes difficult the $\alpha$-FeSe to CuAu-type FeSe transition. We must note that Fig.~\ref{T-path} can give an indication of structural phase transitions of $\alpha$-FeSe under pressure, and the volumes discussed above are the critical volumes at which $\alpha$-FeSe crosses the binding curve of the other phases, \textit{e.g.} NiAs-type FeSe

The above discussion revealed the relative structural stability of $\alpha$-FeSe. However, to have complete understanding of the phase transition of $\alpha$-FeSe, we calculated the transition pressures. The transition pressures ($p_t$) from $\alpha$-FeSe to the other structure, \textit{e.g.}, NiAs-type FeSe, can be obtained either from the slope of the common tangent of both $E$-$V$ curves in Fig.~\ref{T-path} or from the usual condition of equal enthalpies, \textit{i.e.}, the pressure $p_t$ at which the enthalpies $H$ of $\alpha$-FeSe and NiAs-type FeSe are the same. We follow the latter approach, since it is well known that the thermodynamic stable phase at some given pressure and zero temperature is that one with the minimum enthalpy~\cite{prb-enth}. At $p_t$, two phases have the same enthalpy, and then the transition pressure is determined by equating the enthalpies of the two phases of FeSe. Note that this is a standard procedure to calculated the transition pressure of materials~\cite{MgTe,ZnSe,Tin}. To see the possible phase transitions, Fig.~\ref{Pressure} shows the calculated enthalpy curve of $\alpha$-FeSe, NiAs-, and MnP-type FeSe as a function of $p$.

Following the above procedure, the transition pressures of FeSe are calculated and the results are summarized in Table~\ref{ptable}. The GGA (LSDA)  $p_{t}$ of $\alpha$-FeSe to NiAs-type FeSe is calculated to be $\sim 17.5 \; (13.0)$\;{GPa}, whereas $p_t$ of $\alpha$-FeSe to MnP-type FeSe is $\sim 17.70\;(13.05)$\;{GPa} which are in agreement with recent experimental reports~\cite{Medv,Garb}. It is clear that a slight external pressure deforms the hexagonal structure to the orthorhombic one. The experimental reports state that structural phase transition from $\alpha$-FeSe to MnP-type requires $\sim 0.5$\;{GPa} more extra pressure than does NiAs-type~\cite{Medv,Garb}. Our calculations follow the same trend and the GGA (LSDA) results demonstrate that $\sim 0.20 \; (0.05)$\;{GPa} extra pressure than NiAs-type is required for the MnP-type FeSe. We also calulated the transition pressure of $\alpha$-FeSe to CsCl-type FeSe which is $\sim20.45\,${GPa}. Therefore, our calculations reveal that $\alpha$-FeSe takes transition to NiAs-type FeSe, MnP-type FeSe, and CsCl-type FeSe at high pressure irrespective of their magnetic structures. The predictions are in excellent agreement with recent experiments~\cite{Wu2,Medv,Garb}. Table~\ref{ptable} also shows that there is a possibility of a phase transition from NiAs-type to CsCl-type FeSe at $\sim42$ {GPa}. We can see that it is easier for $\alpha$-FeSe to take transition to NiAs- and then  to MnP-type FeSe. However, large external pressure $\sim 60$ {GPa} is required for MnP-type to CsCl-type FeSe phase transition. Further recent experimental observation found CsCl-type FeS~\cite{CsClFeS}, which is isoelectronic to FeSe, at high pressure and this supports the expectation of CsCl-type FeSe~\cite{Rahman2010}.

\subsection{Magnetism}
The calculated magnetic moments per Fe atom ($\mu_\mathrm{Fe}$) are shown in Fig.~\ref{MMGGA}. We did not find any magnetism in $\alpha$-FeSe in the whole volume range, which is in agreement with recent first-principles calculations~\cite{Lee-PRB} and experiment~\cite{Wu2}. The steps in $\mu_\mathrm{Fe}$ of NiAs-type FeSe indicate that the magnetism of FeSe is sensitive to volume compression and expansion. Below the volume of $22$\;{\AA$^3$}, NiAs-type FeSe has zero magnetic moment and the onset of magnetization can be seen for volumes greater than $22$\;{\AA$^3$}; the small step in $\mu_\mathrm{Fe}$ is due to the onset of magnetization. Once the magnetization develops, $\mu_\mathrm{Fe}$ increases slightly and has a constant slope in the volume region $23$--$25$\;{\AA$^3$}; this is the region where MnP-type FeSe is more stable in energy than NiAs-type FeSe.  The step in $\mu_\mathrm{Fe}$ happens at $\sim 31$\;{\AA$^3$}, because near this volume  the hexagonal FeSe is more stable in energy than tetragonal FeSe. Interestingly, similar steps in $\mu_\mathrm{Fe}$ can also be seen for MnP-type FeSe, where the onset of magnetization shifts to a higher volume region as compared with NiAs-type FeSe, \textit{i.e.}, the onset of magnetization starts at $\sim 25$\;{\AA$^3$}, exactly at this volume that the FM MnP-type enters into FM NiAs-type FeSe. Then $\mu_\mathrm{Fe}$ increases with increasing lattice volume and the change in slope can be seen around the volume region where $\alpha$-FeSe crosses the energy curve of MnP-type FeSe. The CuAu-type FeSe has zero $\mu_\mathrm{Fe}$ for volumes $\le 22$\;{\AA$^3$} and $\mu_\mathrm{Fe}$ monotonically increases with the lattice volume. CsCl-type FeSe is more interesting due to its nonzero $\mu_\mathrm{Fe}$ in the whole region of volume and this may indicate that while even all the phases of FeSe become NM, CsCl-type FeSe retains its magnetic structure in an unconventional manner~\cite{Rahman2010}.

\subsection{Electronic Structures}
The phase transitions mentioned are understood by comparing the electronic structures in terms of the density of states (DOS).
The calculated total and atom projected DOS in the FM states are shown in Fig.~\ref{DOS}. The total DOS of $\alpha$-FeSe shows metallic behavior which is in agreement with recent photoemission and transport measurements~\cite{Wu}. The exchange splitting is absent in $\alpha$-FeSe and the Se $p$-states lie well below the Fermi energy ($E_\mathrm{F}$). The Se $p$-states are hybridized with the Fe $d$-states and this hybridization is strong at a lower energy, $\sim -4.0$\;{eV}. The strong peak near $E_\mathrm{F}$ is formed mainly from the Fe $e_{g}$-states, and the Fe $t_{2g}$-states
are more delocalized in energy and extend over a wide range of energy. The DOS shows that $E_\mathrm{F}$ is dominated by the Fe $d$-states, which is also found in recent experimental results on $\alpha$-FeSe~\cite{Yoshida} and the FeAs superconductor~\cite{Sato}. One can also see a pseudogap in the electronic structure of $\alpha$-FeSe which was also found in previous calculations~\cite{singh} and in the FeAs based superconductor~\cite{Fengjie}. This reveals that the electronic
structure of $\alpha$-FeSe is essentially similar to those of FeAs based superconductors.

The NiAs-type FeSe DOS in Fig.~\ref{DOS}(b) shows a metallic property and now there is an exchange splitting between the majority and minority spins, which was absent in $\alpha$-FeSe. Therefore, the structural phase transition is accompanied by a magnetic phase transition. Simultaneously, the Fe local symmetry is also changed from tetrahedral to octahedral. The Se $p$-states extend to lower energies and strongly hybridize mainly with the Fe $t_{2g}$-states. Furthermore, now there is no pseu\-do\-gap in the majority spins, but it exists in the minority spin states. The peak at $\sim -2.0$\;{eV} is contributed by the Fe $t_{2g}$-states. This switch of $e_{g}$ to $t_{2g}$ states is due to the change of local symmetry. Once the local symmetry of the Fe atom is changed to hexagonal (octahedral), further breaking of hexagonal symmetry will not have much affect on the electronic structure as seen in MnP-type FeSe DOS in Fig.~\ref{DOS}(c). The electronic structure is identical to NiAs-type FeSe except for some minor changes around $E_\mathrm{F}$ in the majority and minority spins. In the majority spins, $E_\mathrm{F}$ lies in the valley; in the case of minority spins, the splitting between the $t_{2g}$ and $e_{g}$ states at $E_\mathrm{F}$ is increased and this is probably due to the breaking of hexagonal symmetry.

In the case of NiAs-type FeSe, these minority spin states were nearly degenerate, but the lattice distortion removes this degeneracy and the $e_{g}$-states move to higher energy in MnP-type FeSe, lower the electronic energy and $E_\mathrm{F}$ falls in the valley. One can see that the lattice distortion significantly decreased the DOS at $E_\mathrm{F}$, and the MnP-type FeSe has a lower DOS at $E_\mathrm{F}$ than does NiAs-type FeSe. In the case of CsCl-type FeSe, the majority Fe $e_{g}$-states are completely occupied and the $t_{2g}$-states are partially occupied. Metallic conduction is mainly taken by the $t_{2g}$ spins in the majority as well as in the minority spin states. The total DOS shows the mixing of the Fe-$d$ and Se-$p$ states. The total DOS of CuAu-type FeSe also shows metallic character; the hybridization between the Fe-$d$ and Se-$p$ electrons can be seen. Note that both the $t_{2g}$ and $e_{g}$ spins are completely occupied in the majority states. The exchange splitting between the $t_{2g}$ and $e_{g}$ spins in the majority state is weaker than in the minority spin states. As we saw that our calculated systems are metallic in in low and high pressure regions and the electronic structures agree with the previous experimental and theoretical data which suggest us that the effect of onsite Coulomb interaction, i.e., LDA+U type interaction  in FeSe will not have  significant affects. 

To shed more light on the structural phase transition of $\alpha$-FeSe and to see how the chemical bonding changes, the charge densities were also calculated and plotted in  Fig.~\ref{CDN} in a plane that contains both the Fe and Se atoms. The $\alpha$-FeSe phase forms an enhanced density on the side of the Fe atom. It is clear that the $\alpha$-FeSe bonds are more covalent, a bonding that causes the apparent deviation of the charge density contours from a spherical shape. Inside the Fe-Se block, the metal-like Fe-Fe (Se-Se) bonding occurs due to delocalized Fe $3d$ (Se $p$) states. Interestingly, similar bonding was also found in the FeAs superconductors~\cite{FeP}. Note that such a bonding is either absent or weak in the other crystal structures of FeSe. The structural transition to NiAs-type FeSe also changes the bonding character between the Fe and Se atoms. The charge density of NiAs-type FeSe, see Fig.~\ref{CDN}(b), clearly shows that the bonding between the Fe and Se atoms is less covalent and the charges are more localized at Fe and Se atoms and are nearly spherical. The bonding in MnP-type FeSe is similar to the NiAs-type FeSe. In the CsCl-type FeSe and CuAu-type FeSe, the bonding becomes more ionic and the charge density becomes nearly spherical. It is obvious that the ionic interaction between the Fe-Se bonds becomes stronger and the covalent interaction becomes weaker, when the structural phase transition occurs from $\alpha$-FeSe to the other phases of FeSe. We must note that the role of defects (Se or Fe vacancies) in FeSe can not be ignored and such defects, which are usually in high energy states, may change the magnetic structure of FeSe~\cite{Lee-PRB}. The effect of Se/Fe vacancies on the structural phase transition of FeSe is beyond the scope of the present work.

There might be a relation between superconductivity and covalency because $\alpha$-FeSe is more covalent than the other phases of FeSe, which have not yet shown superconductivity. The ground state structure of FeSe is NM $\alpha$-FeSe and it does not show any magnetic or structural instability at zero pressure. The experimentally observed decrease in the $T_{c}$ of $\alpha$-FeSe at high pressure~\cite{Medv} might be due to a structural/magnetic instability because our calculations show that there is a structural and magnetic instability under high pressure and a structural phase transition takes place to non-superconducting FeSe, \textit{e.g.}, NiAs-type FeSe. Although their is no concensus on the mechanism of the observed superconductivity of $\alpha$-FeSe, the Fermi surface topology plays a crucial role~\cite{Xia} for forming the mediating quasi-particles for Cooper pairs as in cuprates~\cite{JYu}. Once $\alpha$-FeSe transformed into the other structures, such Fermi surface topology and the corresponding van Hove singlularity near the Fermi surface will not be maintained anymore. Therefore, the decrease in $T_{c}$ at high pressure may be due a structural/magnetic instability. This suggests that we may increase the $T_{c}$ of FeSe, if such a structural phase transition at high pressure is avoided. It may be suggested that the structural phase transition in $\alpha$-FeSe can be controlled by doping~\cite{Ding,Serena}. Further work is necessary to control this phase transition or increase the transition pressure of $\alpha$-FeSe.

\section{Summary}
\label{sec:conclusion}
Using \textit{ab initio} calculations, we showed the phase transition of superconducting $\alpha$-FeSe. The predicted structural phase transition and transition pressure were shown to be in excellent agreement with existing experimental observations. Our calculations also showed that there is also a structural phase transition from NiAs-type to MnP-type and CsCl-type FeSe. Phase transition from MnP-type to CsCl-type FeSe requires larger pressure than NiAs-type FeSe. The electronic structure and charge density were similar to the FeAs based superconductors. The calculations revealed that the covalent bond between Fe and Se becomes weaker and the ionic bond become stronger when $\alpha$-FeSe is transformed to other structures. The experimentally observed decrease in $T_{c}$ at high pressure of $\alpha$-FeSe may be due to a structure/magnetic instability. A possible scenario for the enhancement of $T_{c}$ was also discussed.
\\

%\acknowledgements
\noindent\textbf{Acknowledgments}

This work was supported by the Steel Innovation Program by POSCO through POSTECH, 
the Basic Science Research Program (Grant No. 2009-0088216) through the National Research Foundation funded by
the Ministry of Education, Science and Technology of Republic of Korea. The present work was also funded by
the U.\,S.\,N.\,S.\,F (through its MRSEC program at the Northwestern University Materials Research Center).
The authors appreciate Dr. Won Seok Yun for his assistance in figure preparations.

%%%%%%%%%%%%%%%%%%%%%%%%%%%%%%

\newpage
\section*{Tables}
\begin{table}
\caption{Structural parameters of FeSe. The last column shows the optimized volume $V$ 
(in units of \AA$^{3}$) per formula unit} 
\begin{center} 
\begin{tabular}{cccccc}
\hline  \hline
Type & Atom&Atomic positions&$u$ & $v$ & $V$\\
\hline
$\alpha-$FeSe& Fe(2a)& 0 0 0 ; $\frac{1}{2}$  $\frac{1}{2}\,0$&--&--&39.370 \\ 
&Se(2c)& 0 $\frac{1}{2}\, u$ ; $\frac{1}{2}$ 0 ${\bar{u}}$&0.243&--\\ 
\hline
NiAs-type& Fe(2a)&0 0 0; 0 0 $\frac{1}{2}$&--&--&33.640\\
& Se(2b)&$ \frac{1}{3}$ $\frac{2}{3}\, u$ ; $\frac{2}{3}$,$\frac{1}{3}$, $u+\frac{1}{2}$&0.250&--\\
\hline
CuAu-type& Fe(2a)&0 0 0 ; $\frac{1}{2} \frac{1}{2} 0$&--&--&31.890\\
& Se(2b)&$\frac{1}{2}$  0  $\frac{1}{2}$ ; 0  $\frac{1}{2}$  $\frac{1}{2}$&--&--\\
\hline
MnP-type &Fe(4c)&$\pm (u\, v\, \frac{1}{4};\frac{1}{2}-u,v+\frac{1}{2},\frac{1}{4})$ &0.200  &0.005 &29.000\\
&Se(4c)&$\pm (u\, v\, \frac{1}{4};\frac{1}{2}-u,v+\frac{1}{2},\frac{1}{4})$ &0.570  &0.190 \\ 
\hline
CsCl-type& Fe$\;\;\;\;\;\;$& 0 0 0 &--&--&28.370 \\ 
&Se$\;\;\;\;\;\;$&  $\frac{1}{2}\,\frac{1}{2}\,\frac{1}{2}$ &--&--&\\ 
\hline\hline
\end{tabular}
%\end{ruledtabular}
\label{Atomic}
\end{center}
\end{table}

\begin{table}
\caption{Calculate structural transition pressures (in units of GPa) of FeSe. The available experimental data are taken from Ref.~\cite{Garb,Medv}.} 
\label{ptable}
\begin{center}
\begin{tabular}{cccccccccccc}
\hline \hline
&&Theory & Experiment\\
\hline
$\alpha-$FeSe$\rightarrow$& NiAs-type &    17.50  &     12.0 \\      
& MnP-type &    17.70  &     12.5 \\     
& CsCl-type &    20.45  &     $-$ \\ 
NiAs-type$\rightarrow$ &MnP-type &    $21\;\;\;\;\;$  &     $-$ \\ 
&CsCl-type &    $42\;\;\;\;\;$  &     $-$ \\ 
MnP-type$\rightarrow$ &CsCl-type &    $60\;\;\;\;\;$  &     $-$ \\ 
\hline\hline
\end{tabular}
\end{center}
\end{table}
\newpage

\newpage
\section*{Figure captions}
\begin{figure}
\caption{(Color online) Crystal structures of FeSe.(a) $\alpha$-FeSe, (b) NiAs-, (c) MnP-, (d) CsCl-, and (e) CuAu-type FeSe. Red and yellow balls show the Fe and Se atoms, respectively.}
\label{structures}
\end{figure}
 
\begin{figure}
\caption{(Color online) The calculated total energy in units of {eV/f.u.} as a function of lattice volume (\AA$^{3}$) for $\alpha$-FeSe (a) NiAs-, (b) MnP-, (c) CsCl-, and (d) CuAu-type FeSe. The total energy was measured with respect to the total energy of $\alpha$-FeSe equilibrium volume. Filled circles (squares) show the total energy in the NM (FM) state. Red, black, cyan, green, and blue colors represent $\alpha$-FeSe, NiAs-, MnP-, CsCl-, and CuAu-type FeSe, respectively. The inset (e) shows the E---V curve of all the structure in volume interval $20$--$35$\;{\AA$^{3}$}.}
\label{T-path}
\end{figure}

\begin{figure}
\caption{(Color online) The calculated enthalpy in units of {Ry} as a function of pressure (GPa) for (a) $\alpha$-FeSe $\rightarrow$ NiAs-type, (b) $\alpha$-FeSe $\rightarrow$ MnP-type, (c) $\alpha$-FeSe $\rightarrow$ CsCl-type, (d) NiAs-type $\rightarrow$ CsCl-type, (e)NiAs-type$\rightarrow$ MnP-type, and (f) MnP-type $\rightarrow$ CsCl-type FeSe. Red, black, cyan, and green colors represent $\alpha$-FeSe, NiAs-, MnP-FeSe, and CsCl-type FeSe respectively. Dotted lines represent the NM states}
\label{Pressure}
\end{figure}

\begin{figure}
\caption{(Color online) Calculated magnetic moments (in units of $\mu_\mathrm{B}$) per Fe atom as a function
of lattice volume (in units of \AA$^{3}$) of FM FeSe. Red, black, cyan, green, and blue colors represent $\alpha$-FeSe, NiAs-, MnP-, CsCl-, and CuAu-type FeSe, respectively.}
\label{MMGGA}
\end{figure}

\begin{figure}
\caption{(Color online) Total and atom-projected DOS in the FM states of (a) $\alpha$-FeSe, (b) NiAs-, (c) MnP-, (d) CsCl-, and (e) CuAu-type FeSe at their equilibrium lattice volumes. Red, blue, and green lines represent Fe $t_{2g}$, Fe $e_{g}$ and Se $p$ states, respectively, whereas black lines in the bottom panels represent the total DOS. The Se $p$ states are multiplied
by a factor of 10 and the Fermi levels ($E_\mathrm{F}$) are set to zero. The total DOS is given per formula unit for comparison purpose.}
\label{DOS}
\end{figure}

\begin{figure}
\caption{(Color online) Charge-density contours of (a) $\alpha$-FeSe,(b) NiAs-, (c) MnP-, (d) CsCl-, and (e) CuAu-type FeSe in the planes containing Fe and Se atoms. The lowest contour starts from $2 \times 10^{-4}$\;{electrons/a.u.$^3$}
and the subsequent lines differ by a factor $\sqrt{2}$. The color code of the charge density contours is ordered as green, yellow, and red, in ascending order. The dotted lines show the bond between Fe and Se atoms.}
\label{CDN}
\end{figure}
\newpage

\newpage
\begin{center}
\vfill
%\graphics[width=0.8\textwidth]{Fig1a_alpha-FeSe}
\includegraphics[width=0.8\textwidth]{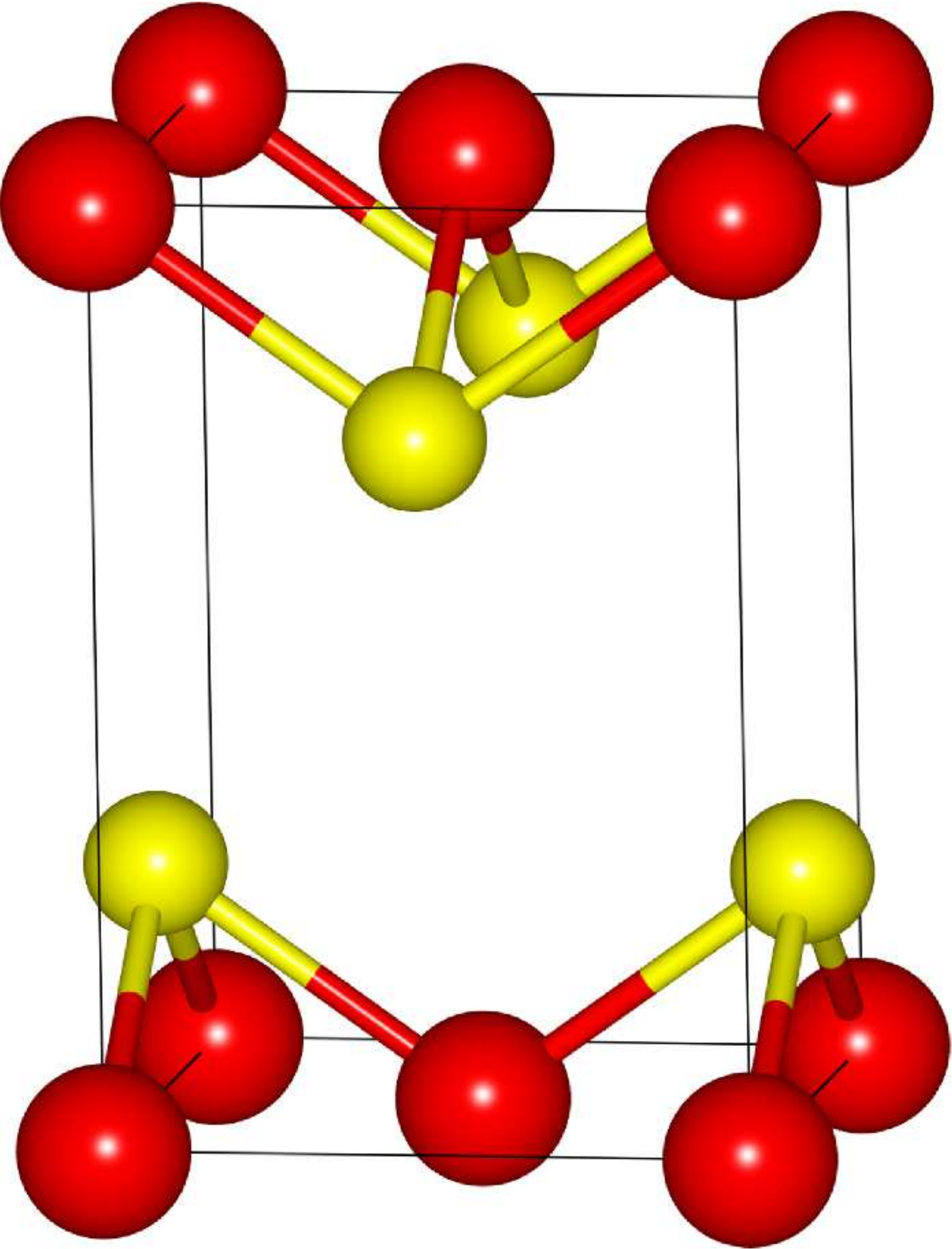}
\vfill
\textbf{Fig. 1(a)}
\end{center}

\newpage
\begin{center}
\vfill
\includegraphics[width=0.8\textwidth]{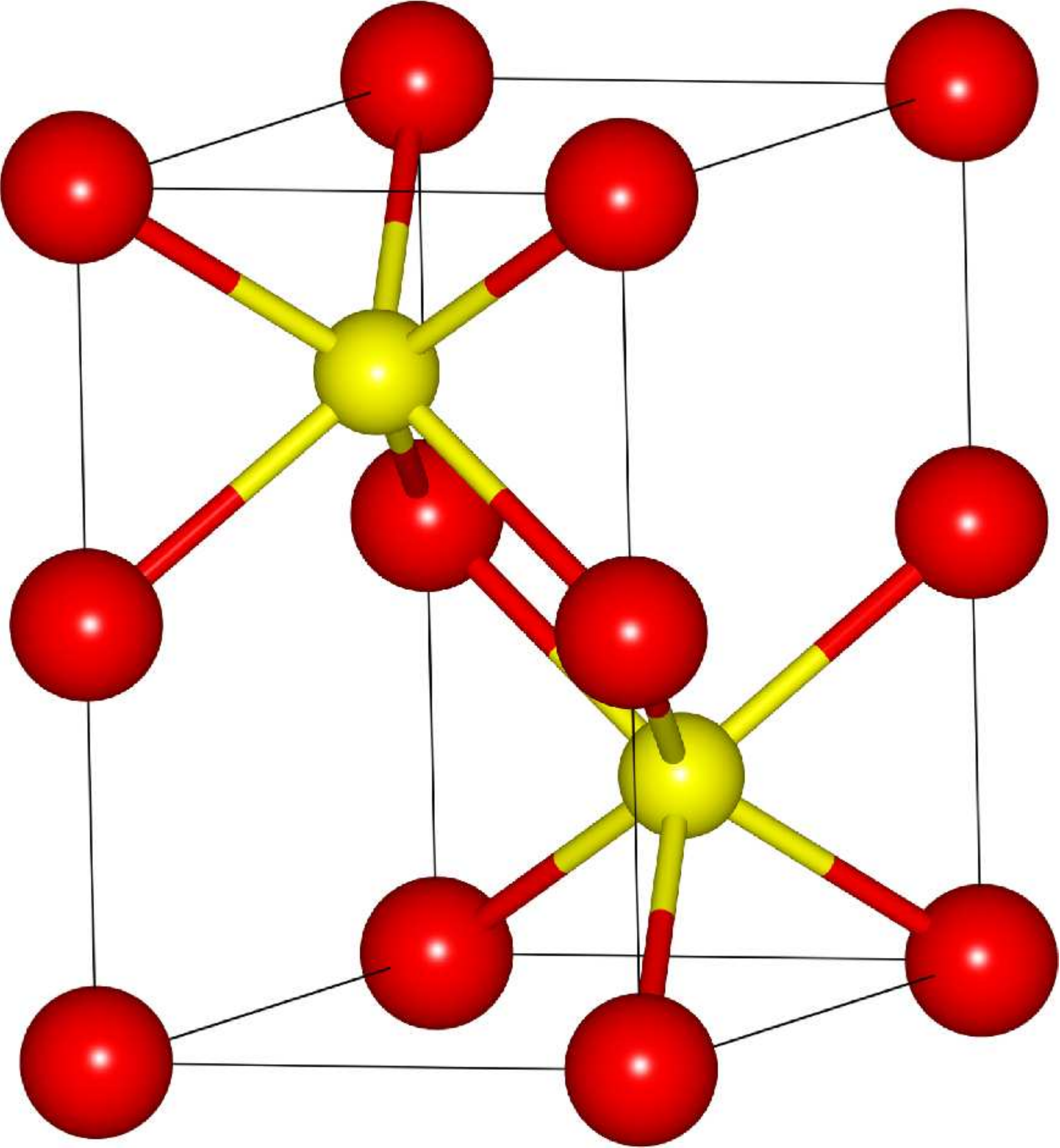}
\vfill
\textbf{Fig. 1(b)}
\end{center}

\newpage
\begin{center}
\vfill
\includegraphics[width=0.8\textwidth]{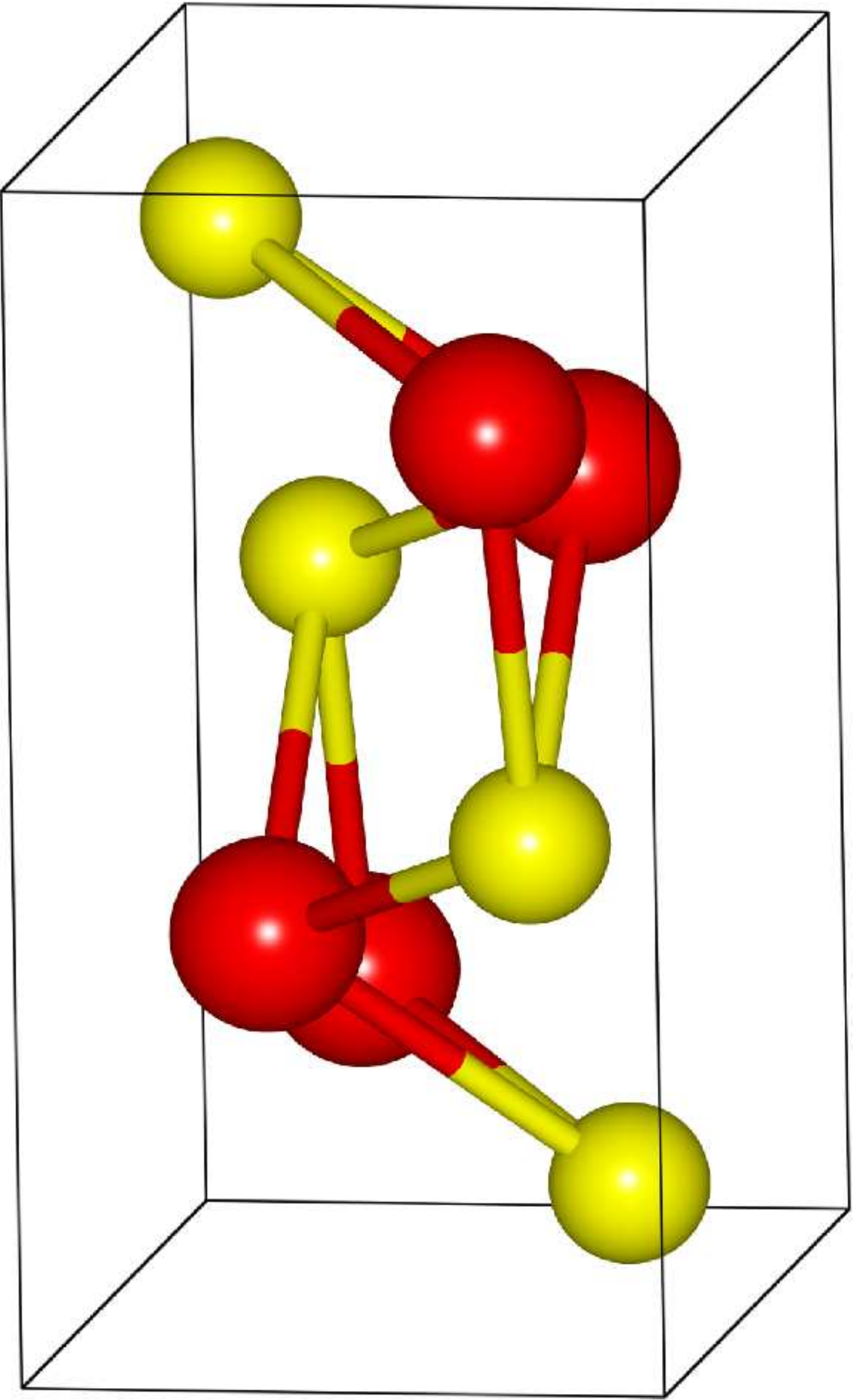}\\
\vfill
\textbf{Fig. 1(c)}
\end{center}

\newpage
\begin{center}
\vfill
\includegraphics[width=0.8\textwidth]{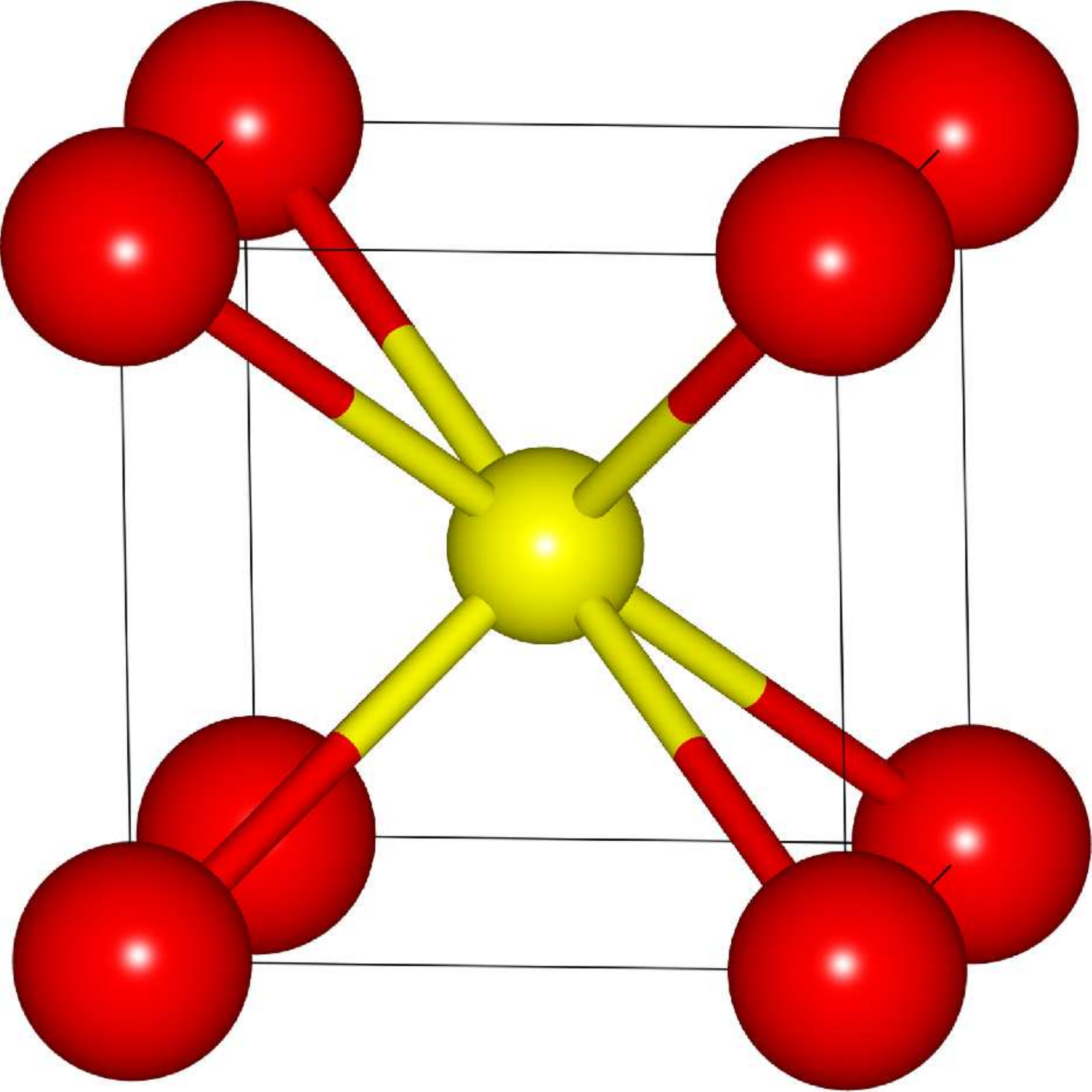}\\
\vfill
\textbf{Fig. 1(d)}
\end{center}

\newpage
\begin{center}
\vfill
\includegraphics[width=0.8\textwidth]{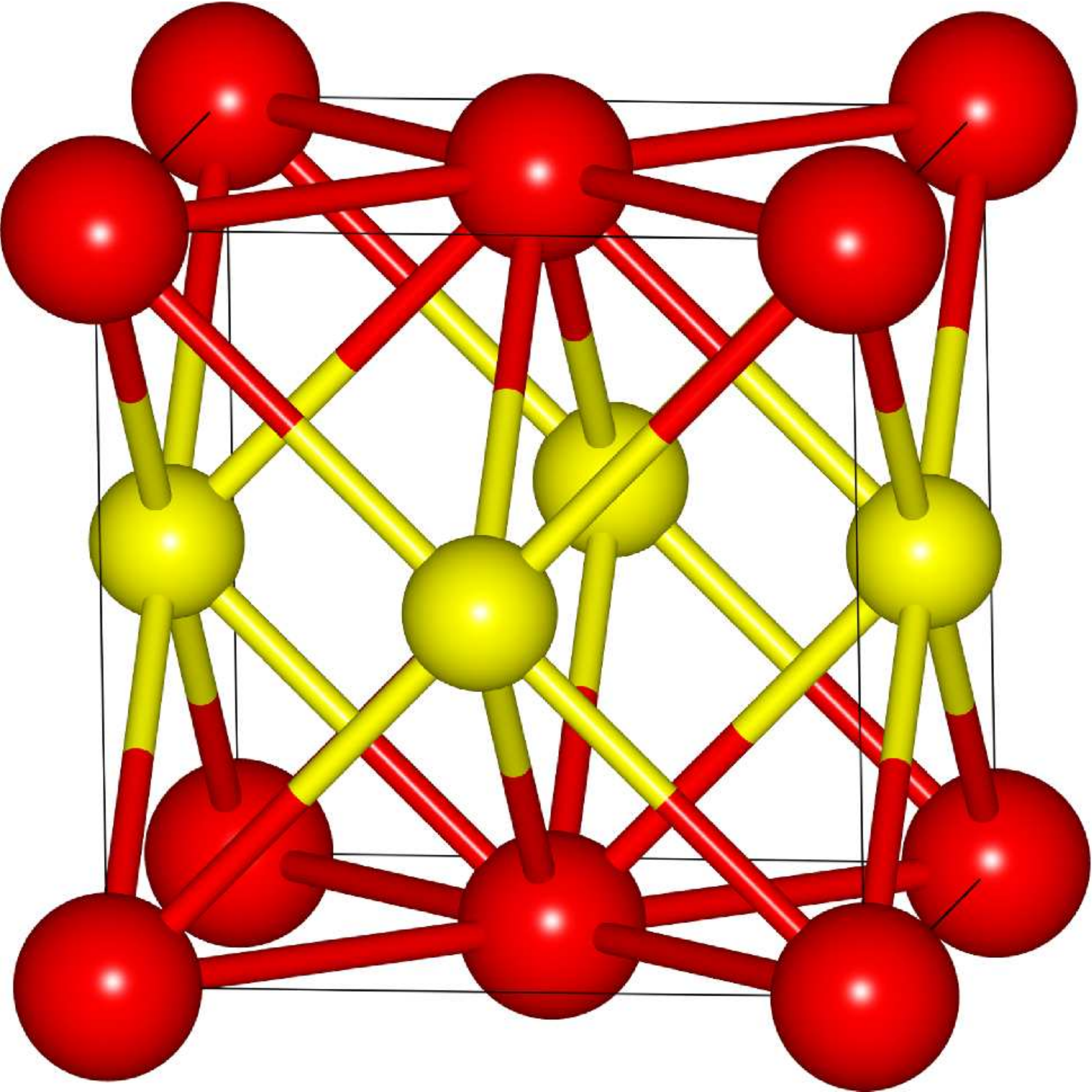}\\
\vfill
\textbf{Fig. 1(e)}
\end{center}

\newpage
\begin{center}
\vfill
\includegraphics[width=\textwidth]{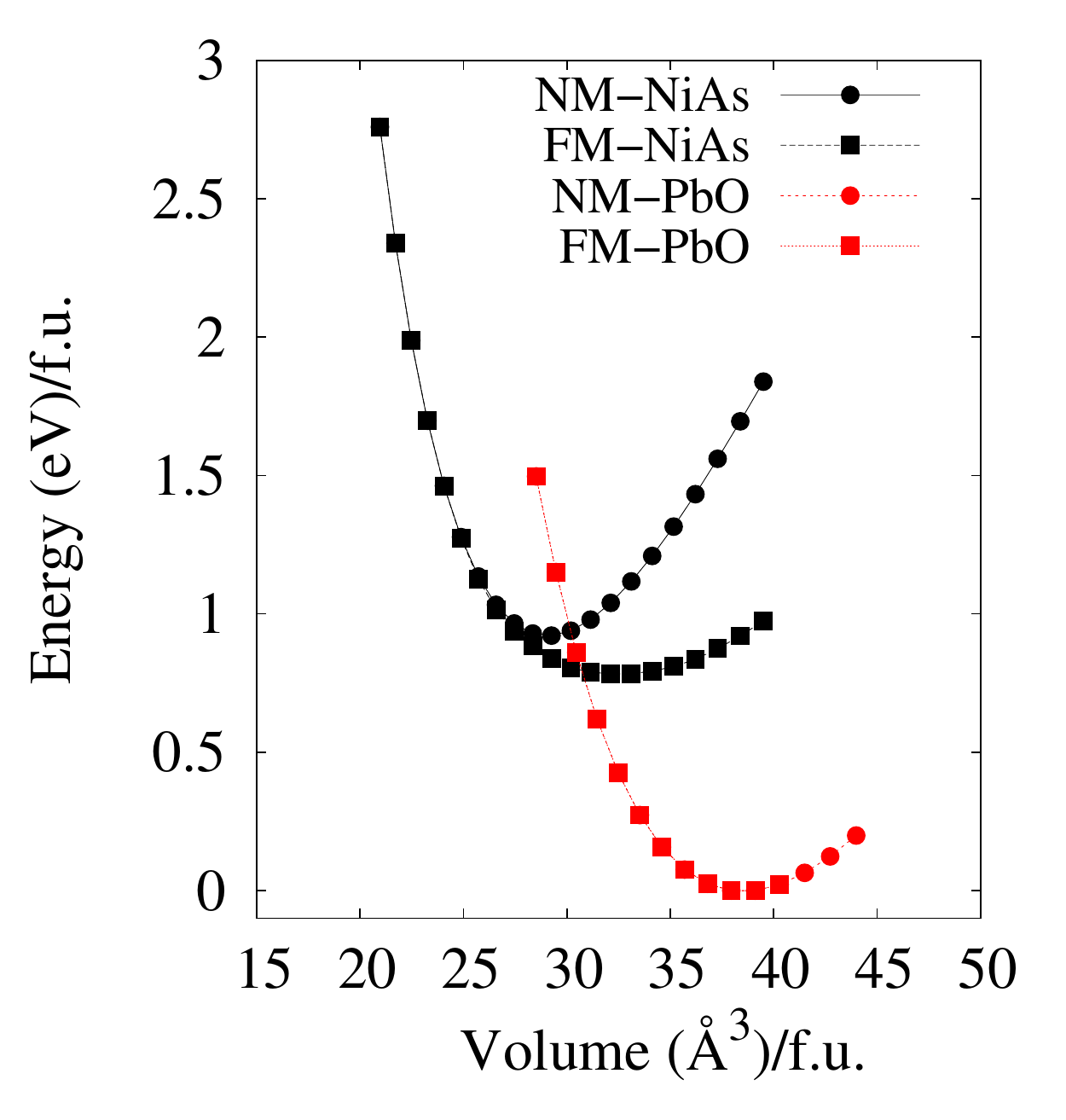}\\
\vfill
\textbf{Fig. 2(a)}
\end{center}

\newpage
\begin{center}
\vfill
\includegraphics[width=\textwidth]{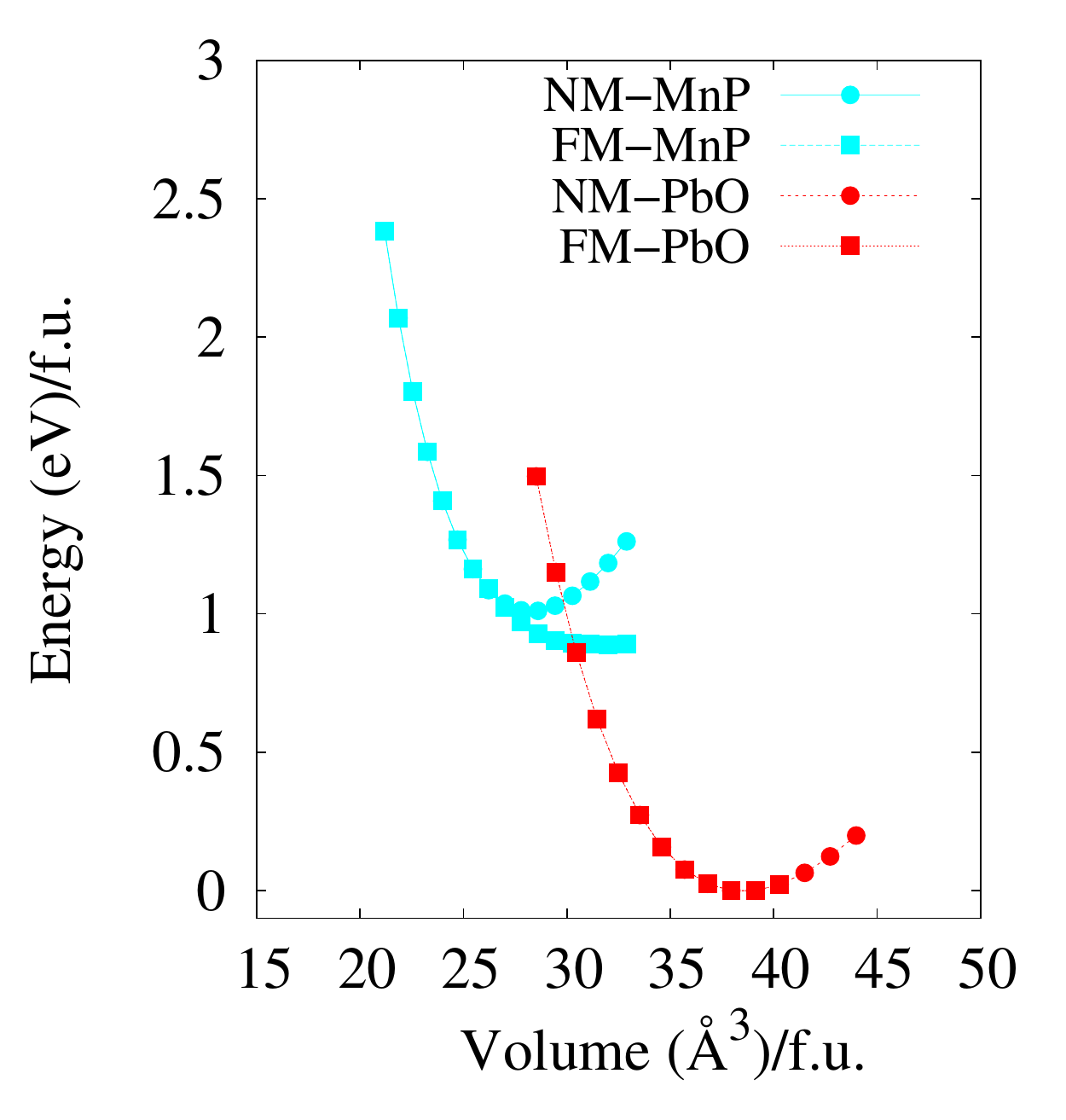}\\
\vfill
\textbf{Fig. 2(b)}
\end{center}

\newpage
\begin{center}
\vfill
\includegraphics[width=\textwidth]{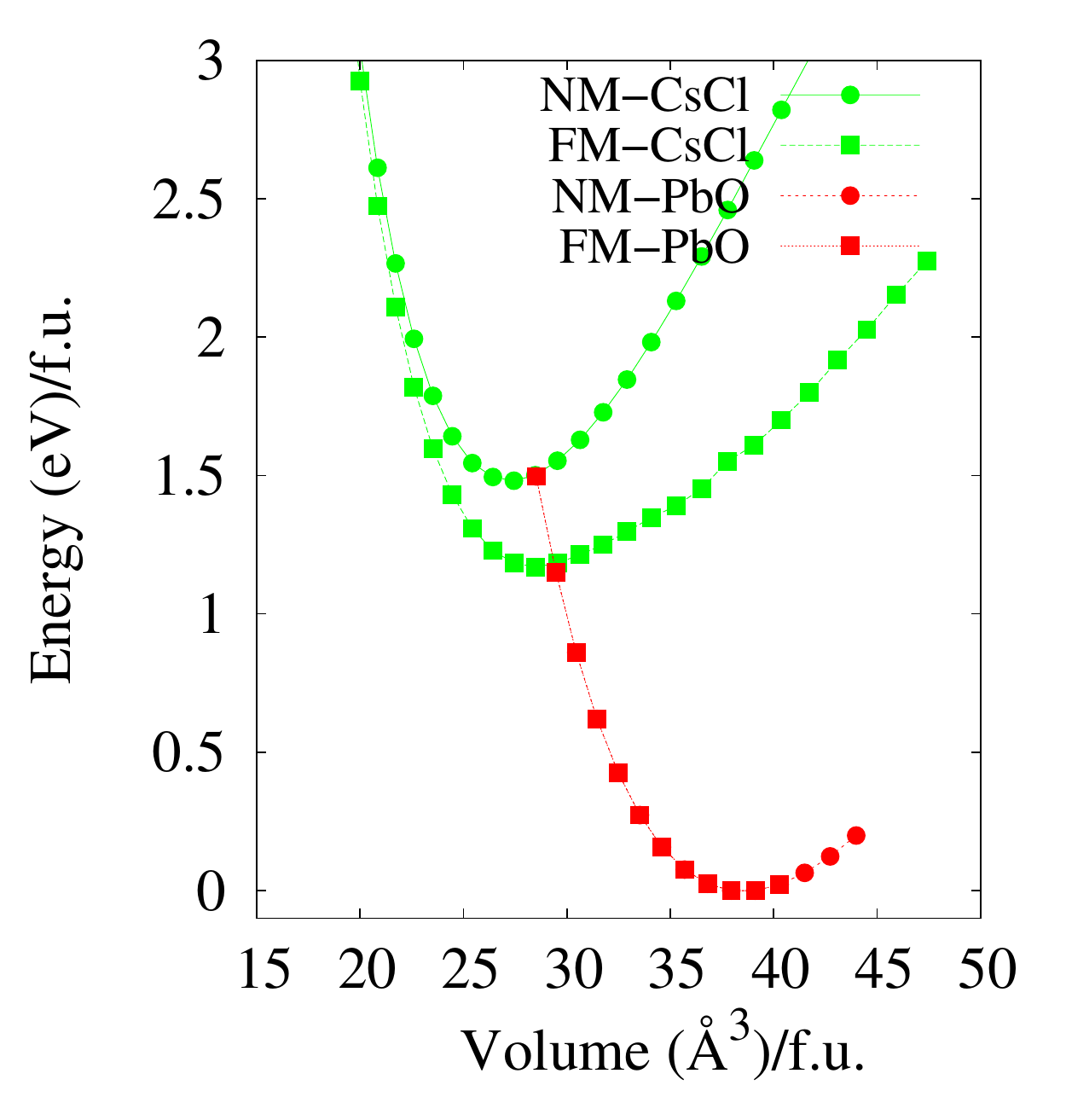}\\
\vfill
\textbf{Fig. 2(c)}
\end{center}

\newpage
\begin{center}
\vfill
\includegraphics[width=\textwidth]{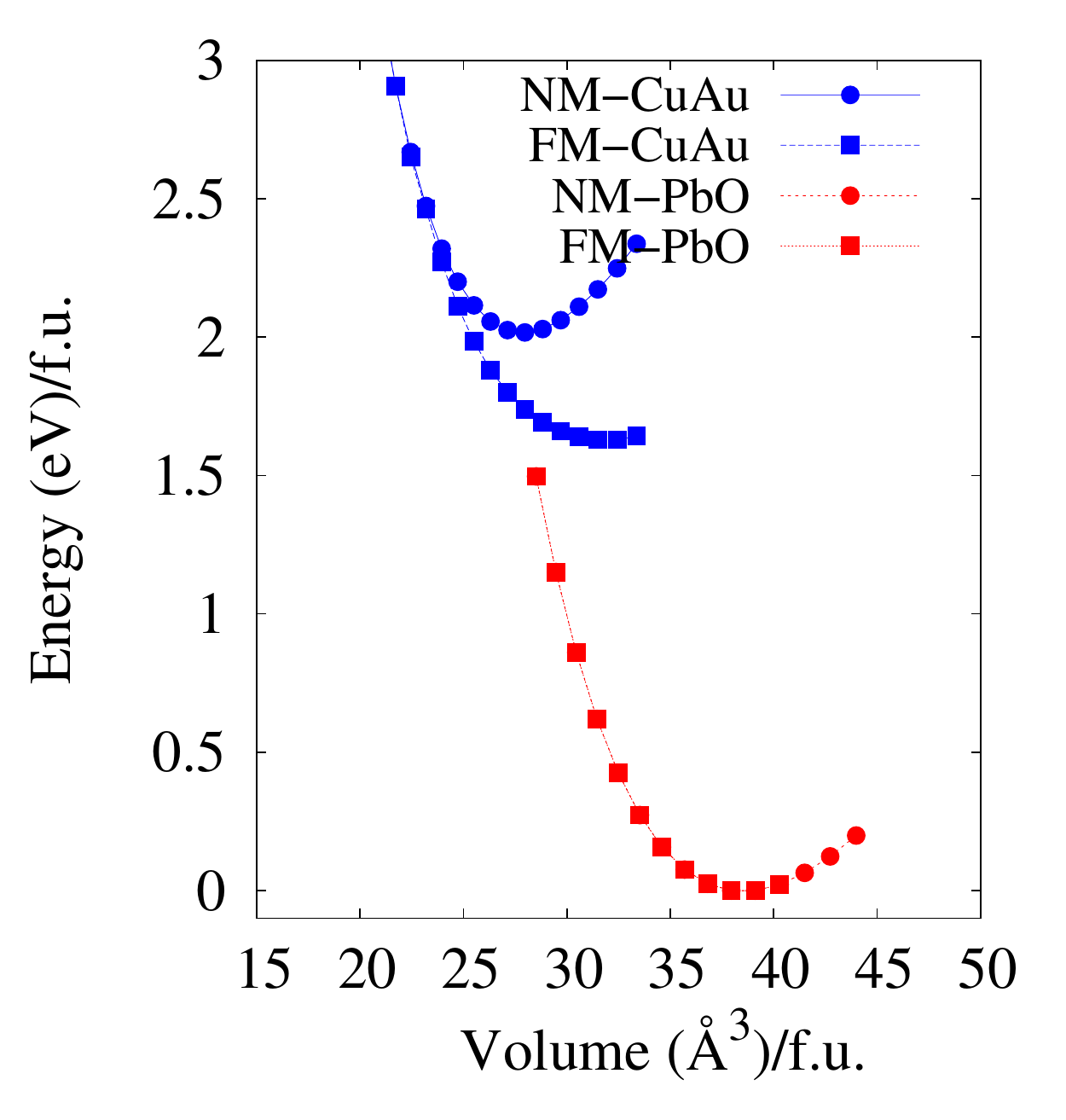}\\
\vfill
\textbf{Fig. 2(d)}
\end{center}

\newpage
\begin{center}
\vfill
\includegraphics[width=0.6\textwidth]{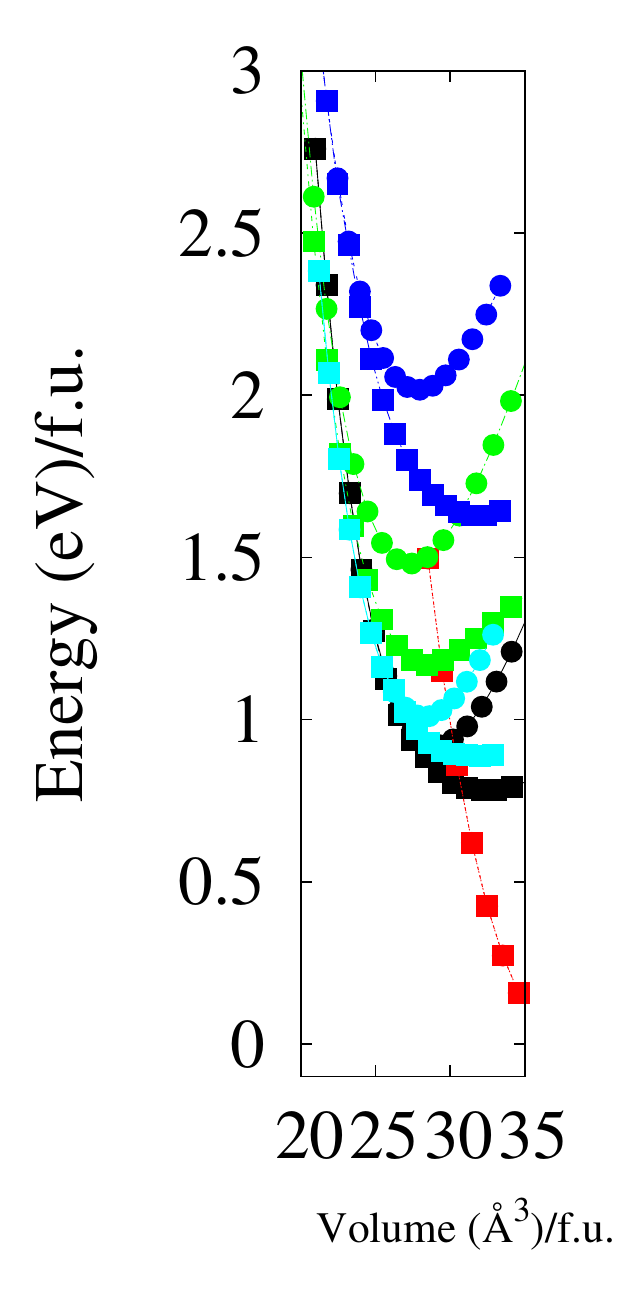}\\
\vfill
\textbf{Fig. 2(e)}
\end{center}

\newpage
\begin{center}
\vfill
\includegraphics[width=0.9\textwidth]{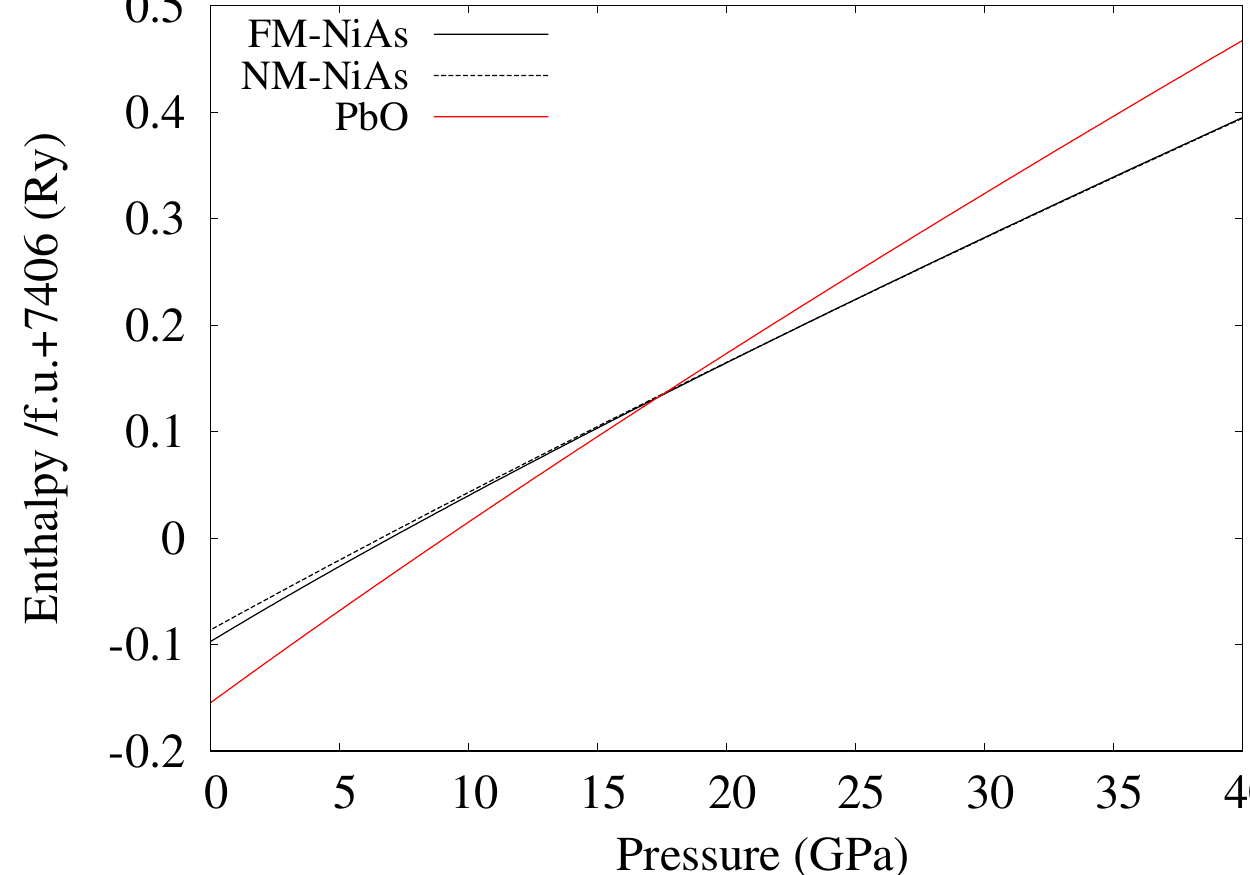}\\
\vfill
\textbf{Fig. 3(a)}
\end{center}

\newpage
\begin{center}
\vfill
\includegraphics[width=0.9\textwidth]{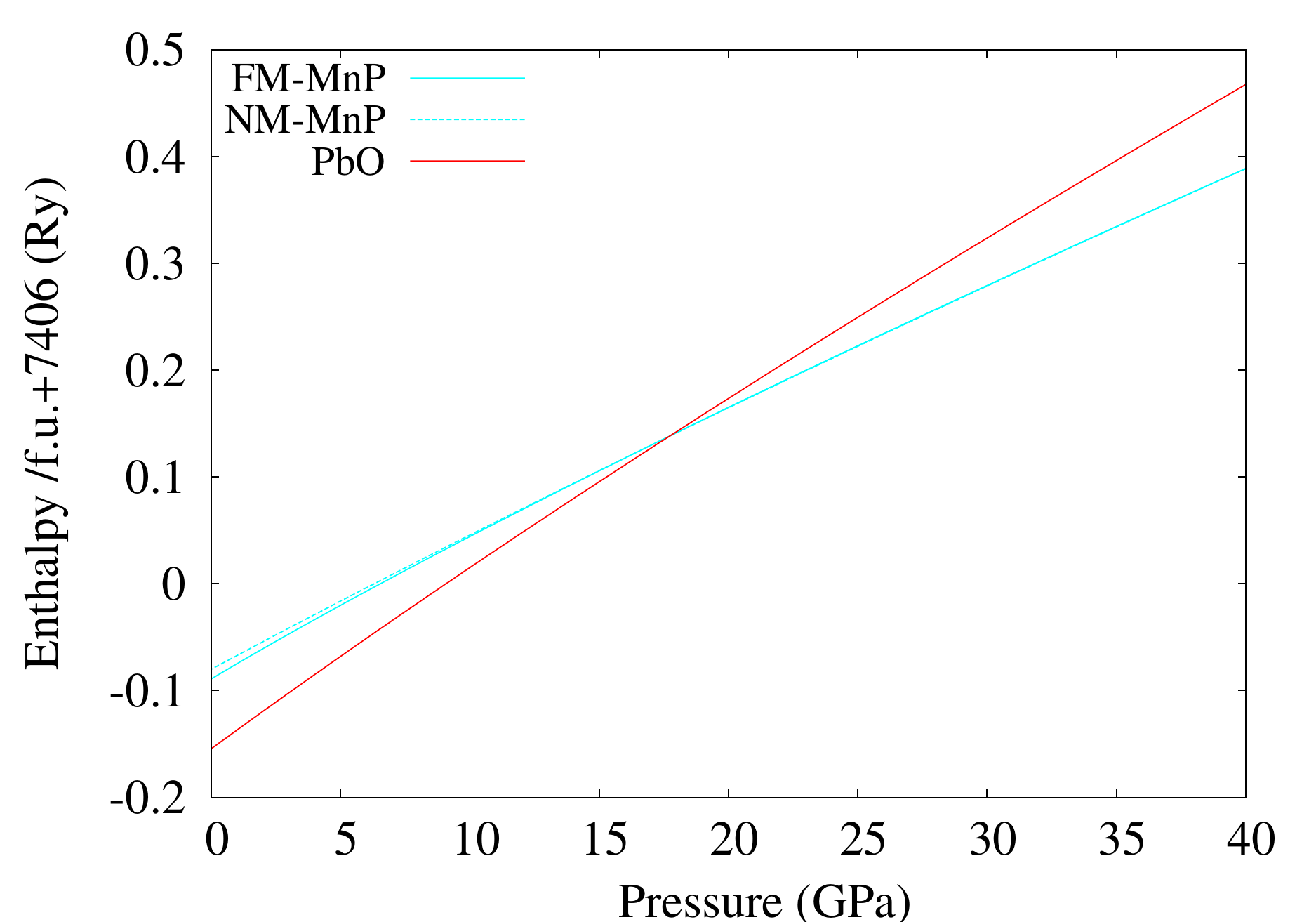}\\
\vfill
\textbf{Fig. 3(b)}
\end{center}

\newpage
\begin{center}
\vfill
\includegraphics[width=0.9\textwidth]{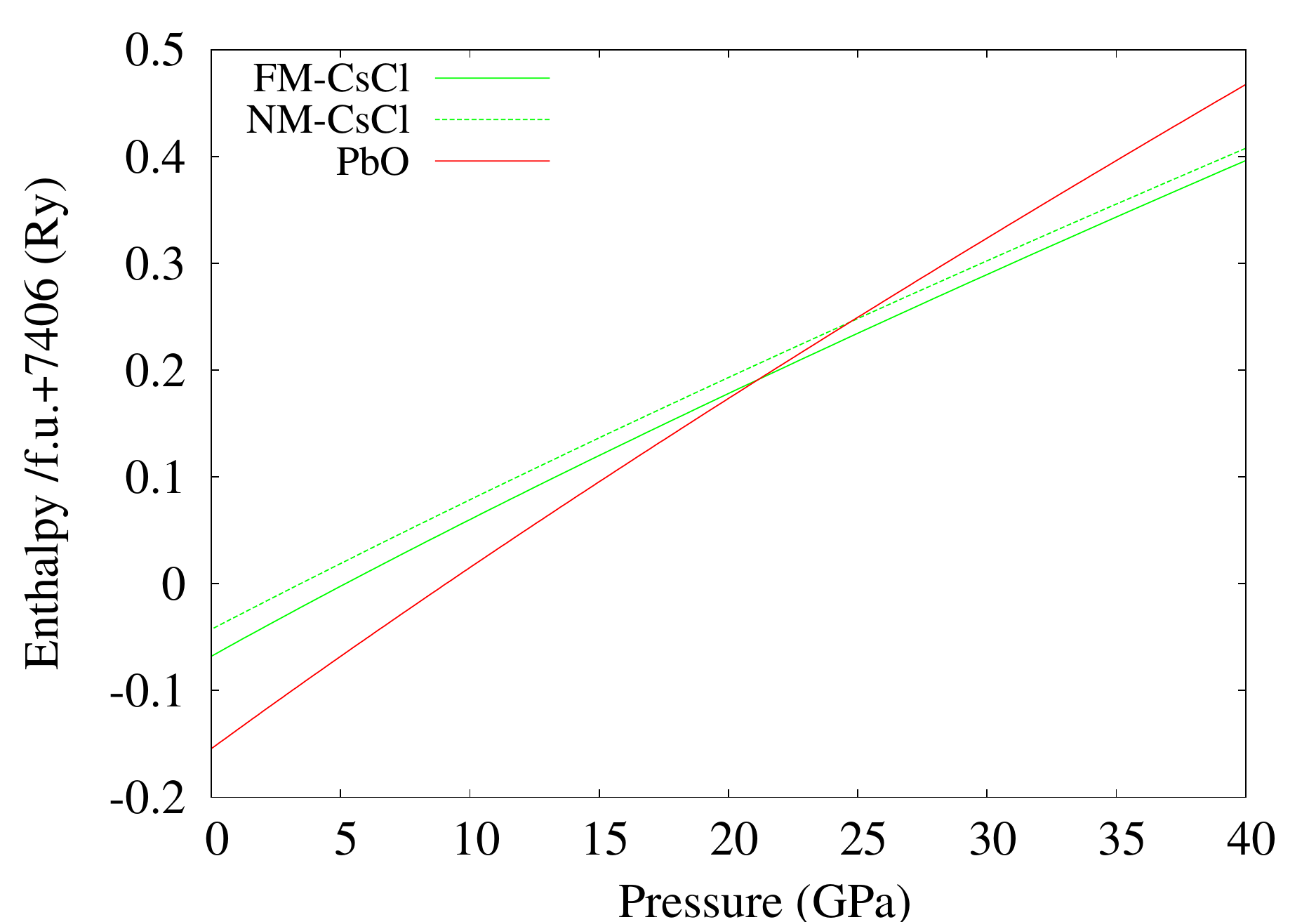}\\
\vfill
\textbf{Fig. 3(c)}
\end{center}

\newpage
\begin{center}
\vfill
\includegraphics[width=0.9\textwidth]{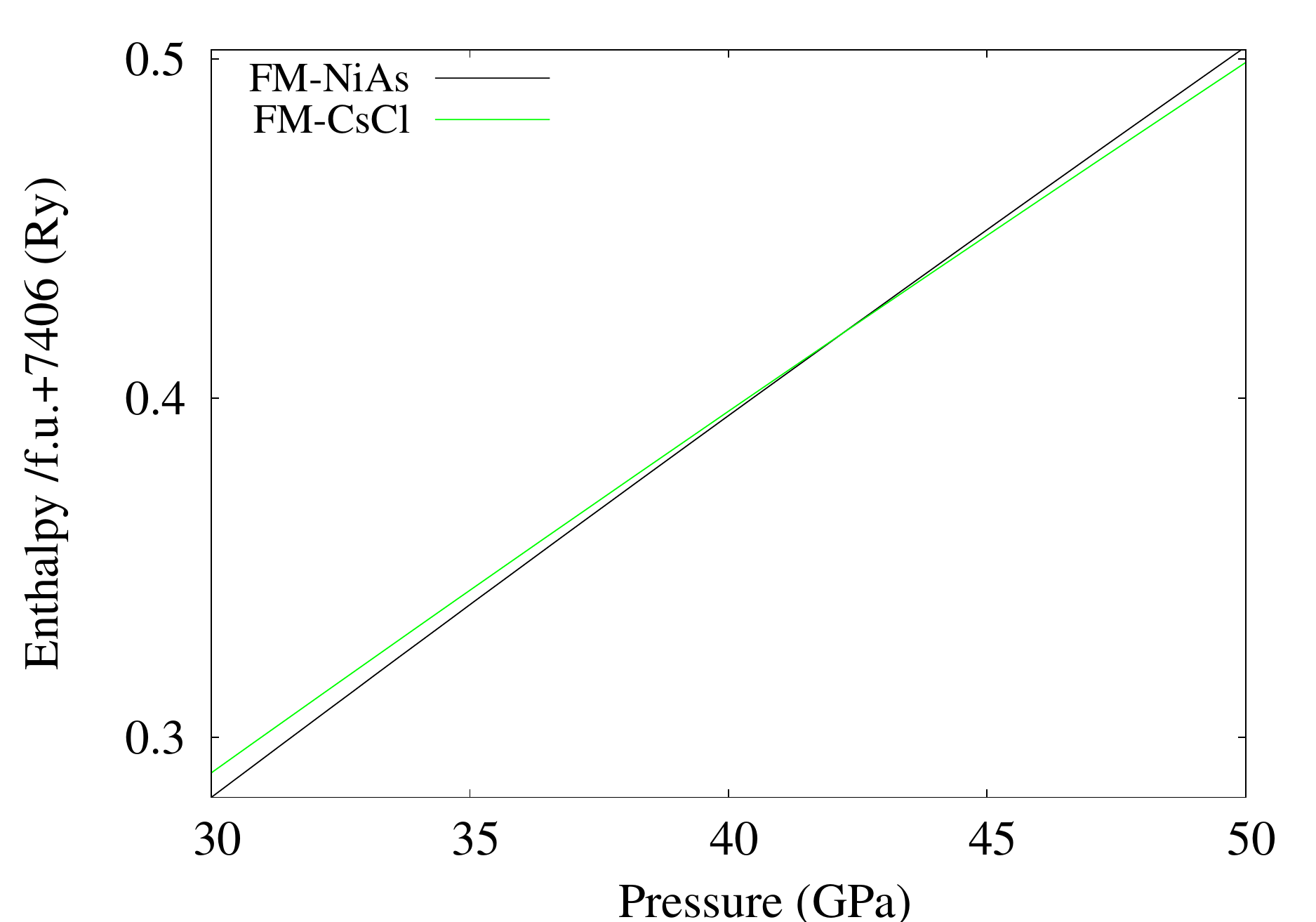}\\
\vfill
\textbf{Fig. 3(d)}
\end{center}

\newpage
\begin{center}
\vfill
\includegraphics[width=0.9\textwidth]{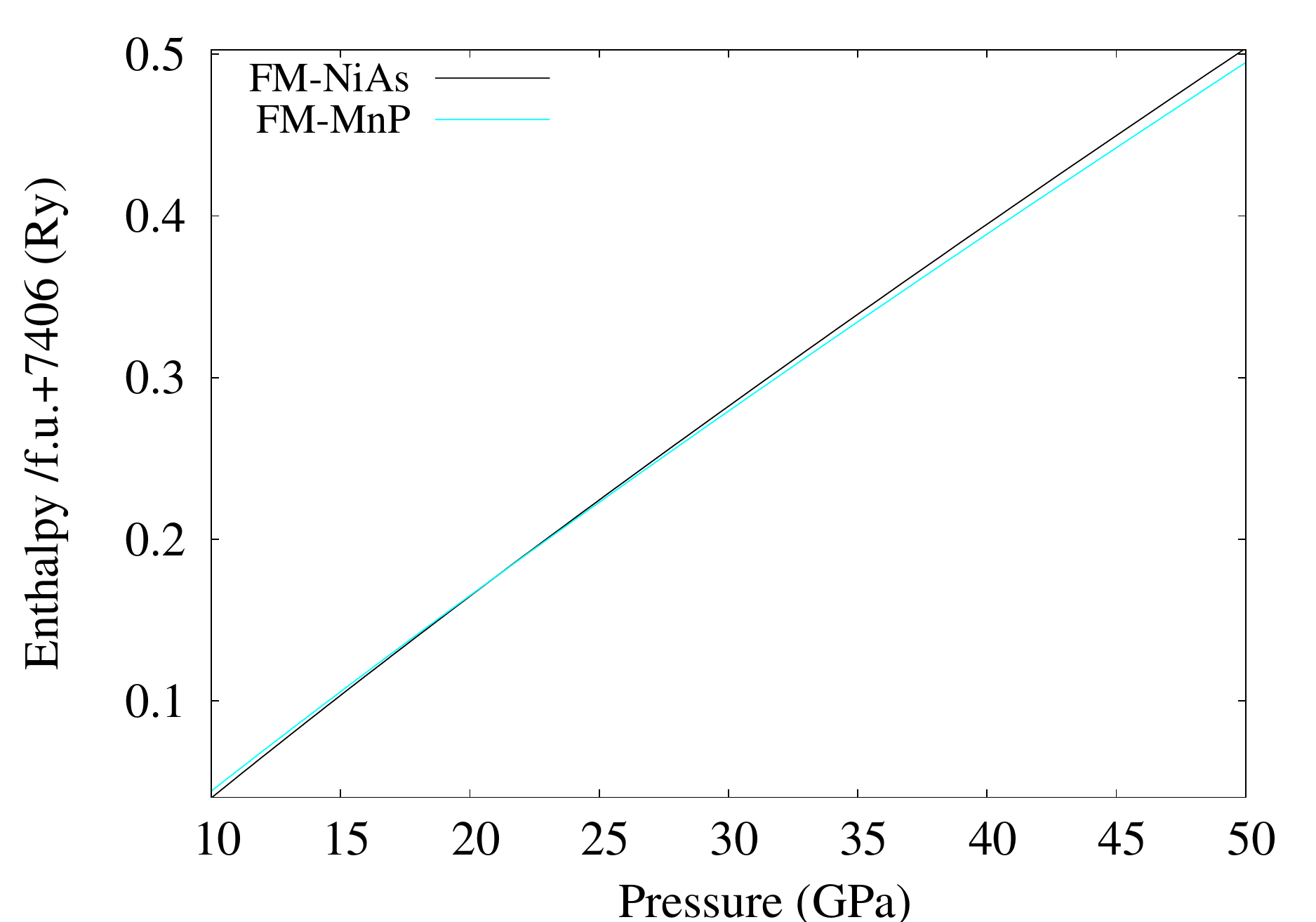}\\
\vfill
\textbf{Fig. 3(e)}
\end{center}

\newpage
\begin{center}
\vfill
\includegraphics[width=0.9\textwidth]{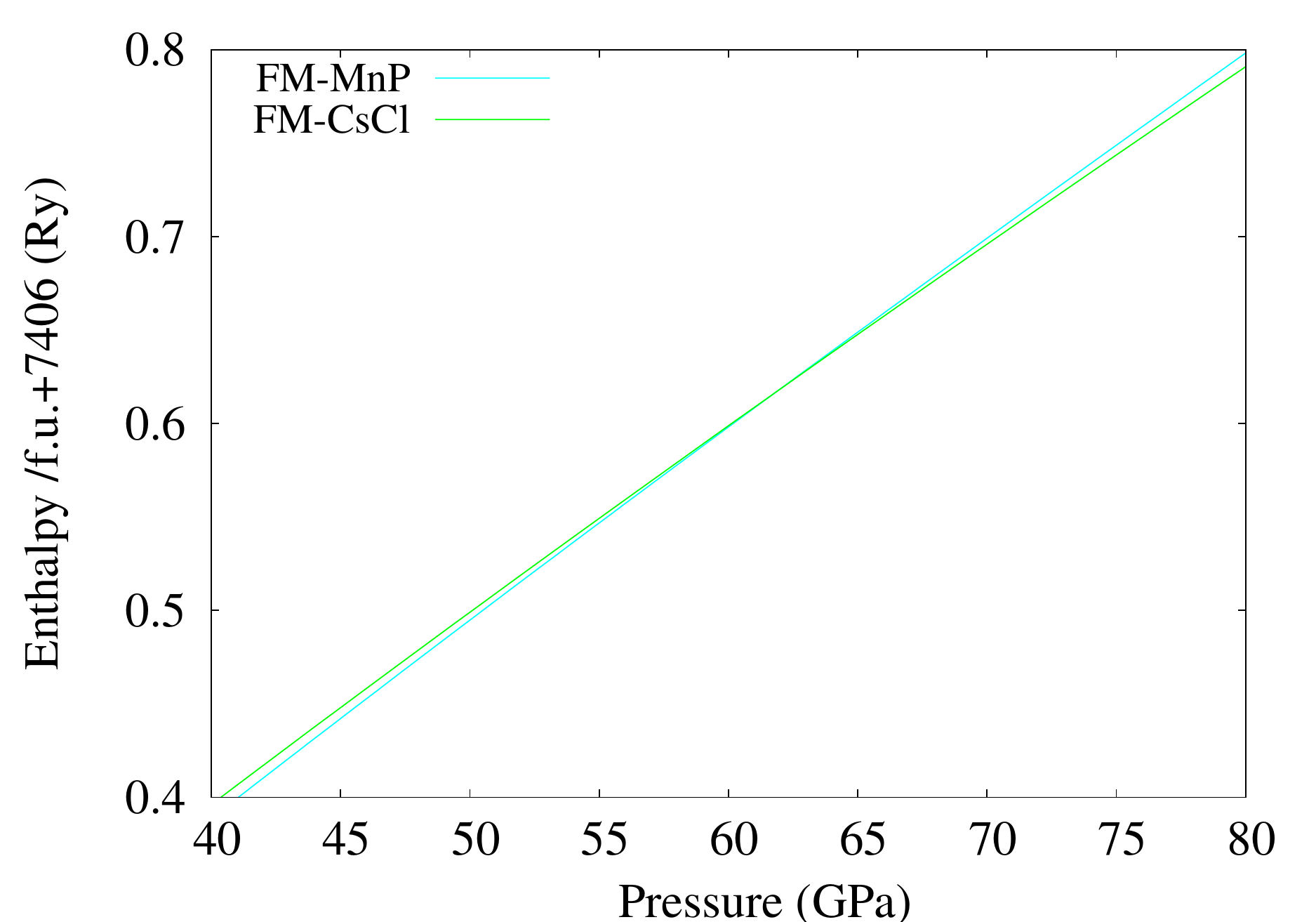}\\
\vfill
\textbf{Fig. 3(f)}
\end{center}

\newpage
\begin{center}
\vfill
\includegraphics[width=\textwidth]{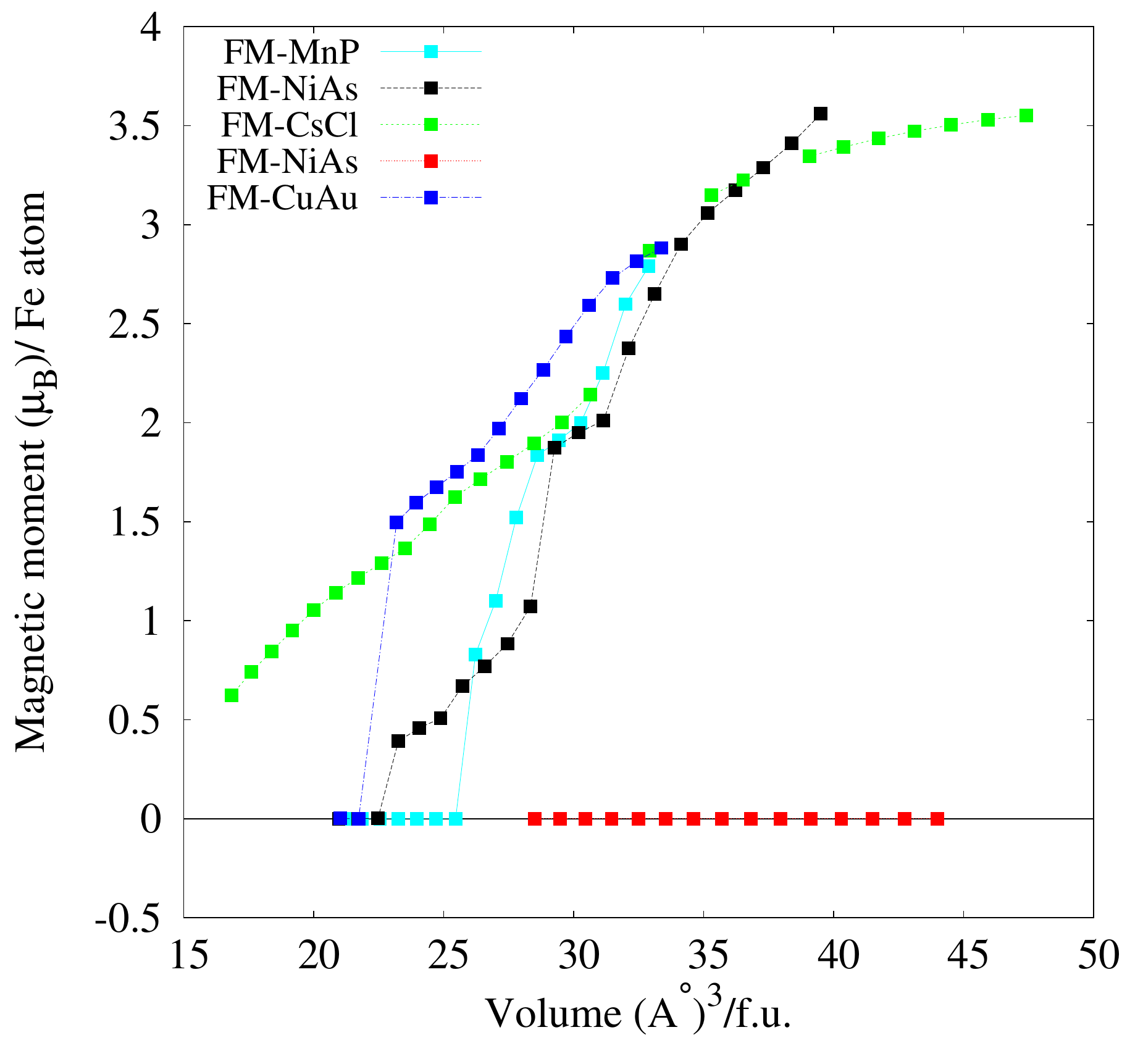}\\
\vfill
\textbf{Fig. 4}
\end{center}

\newpage
\begin{center}
\vfill
\includegraphics[width=0.8\textwidth]{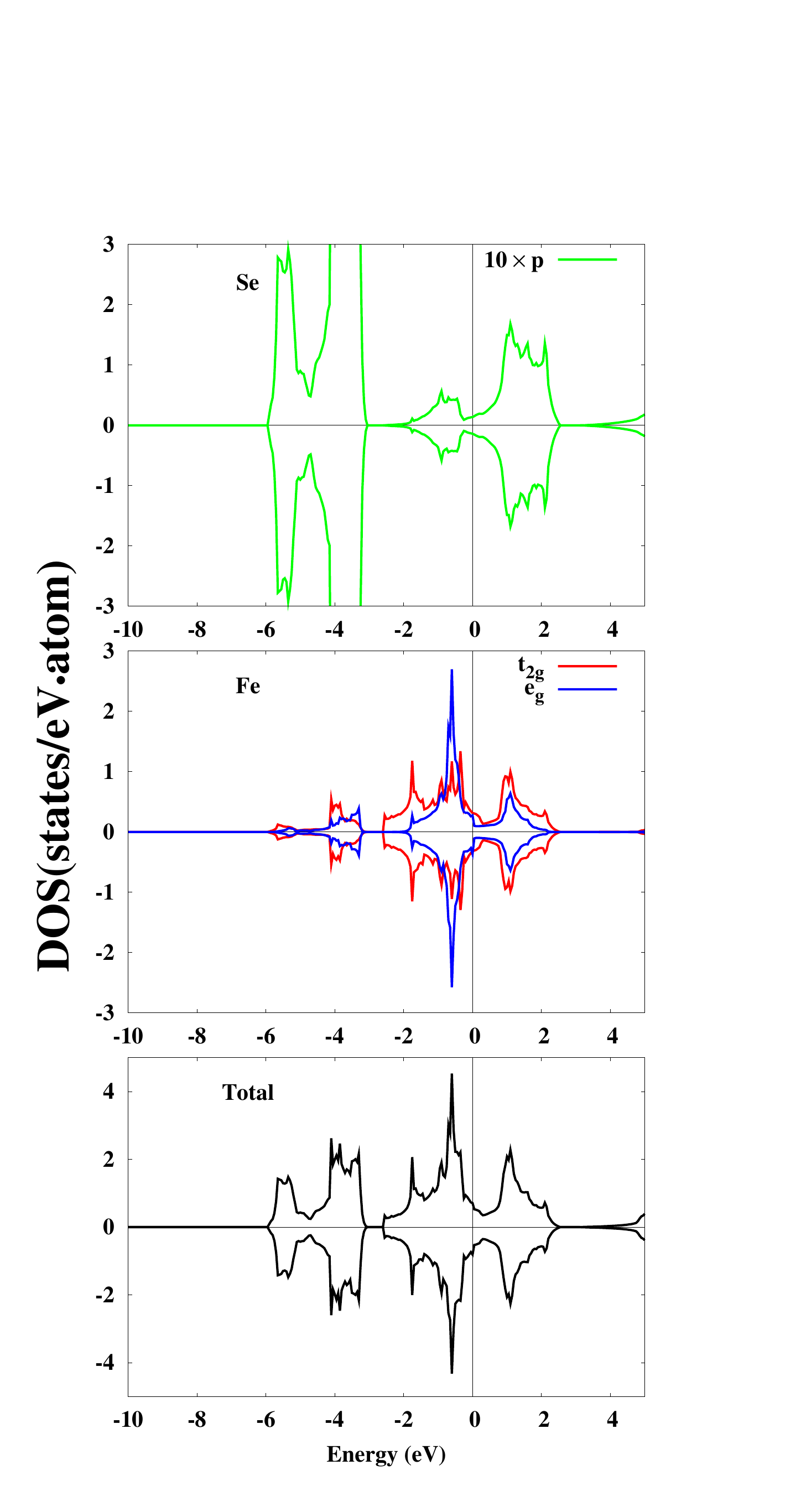}\\
\vfill
\textbf{Fig. 5(a)}
\end{center}

\newpage
\begin{center}
\vfill
\includegraphics[width=0.8\textwidth]{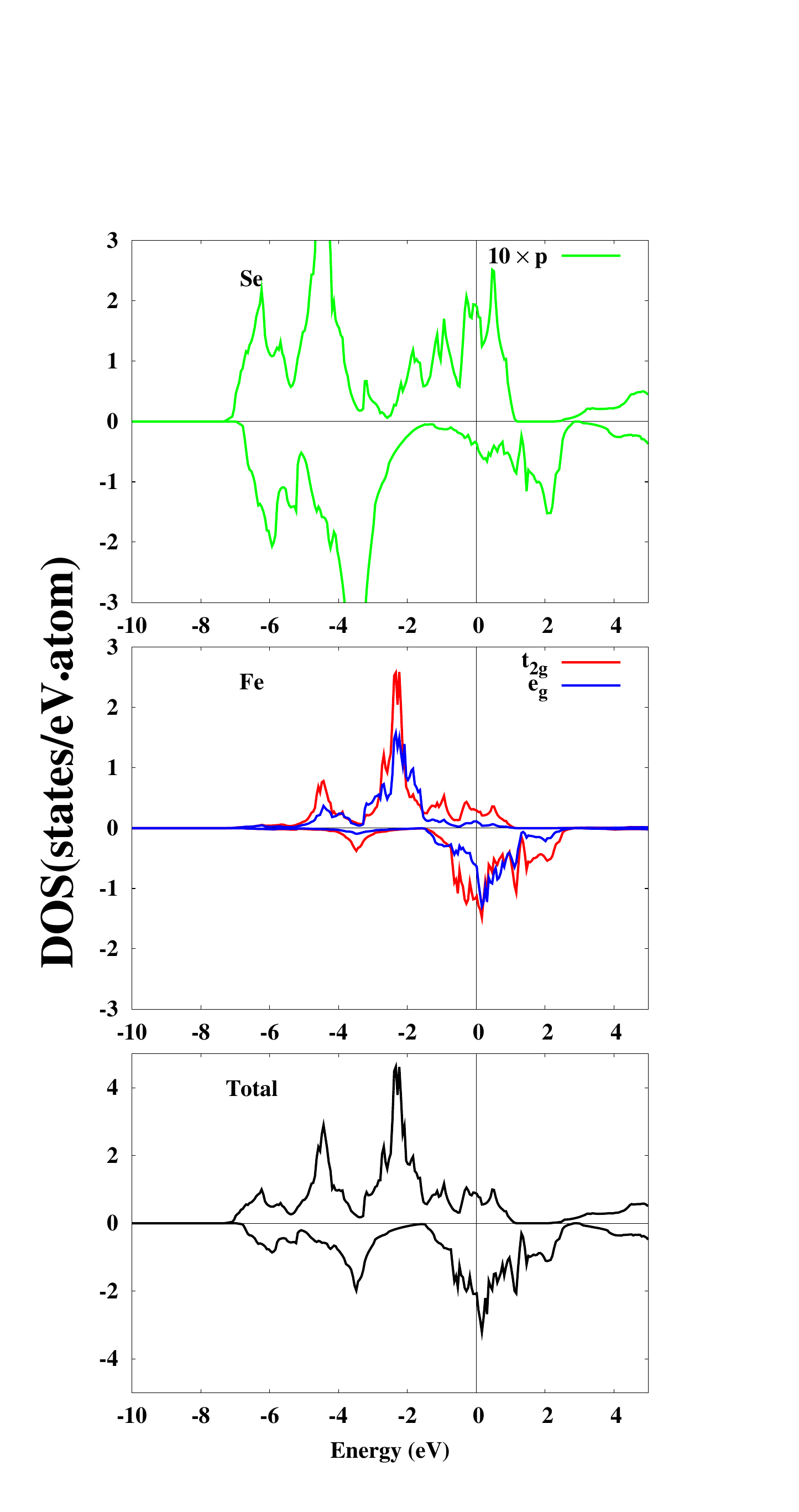}\\
\vfill
\textbf{Fig. 5(b)}
\end{center}

\newpage
\begin{center}
\vfill
\includegraphics[width=0.8\textwidth]{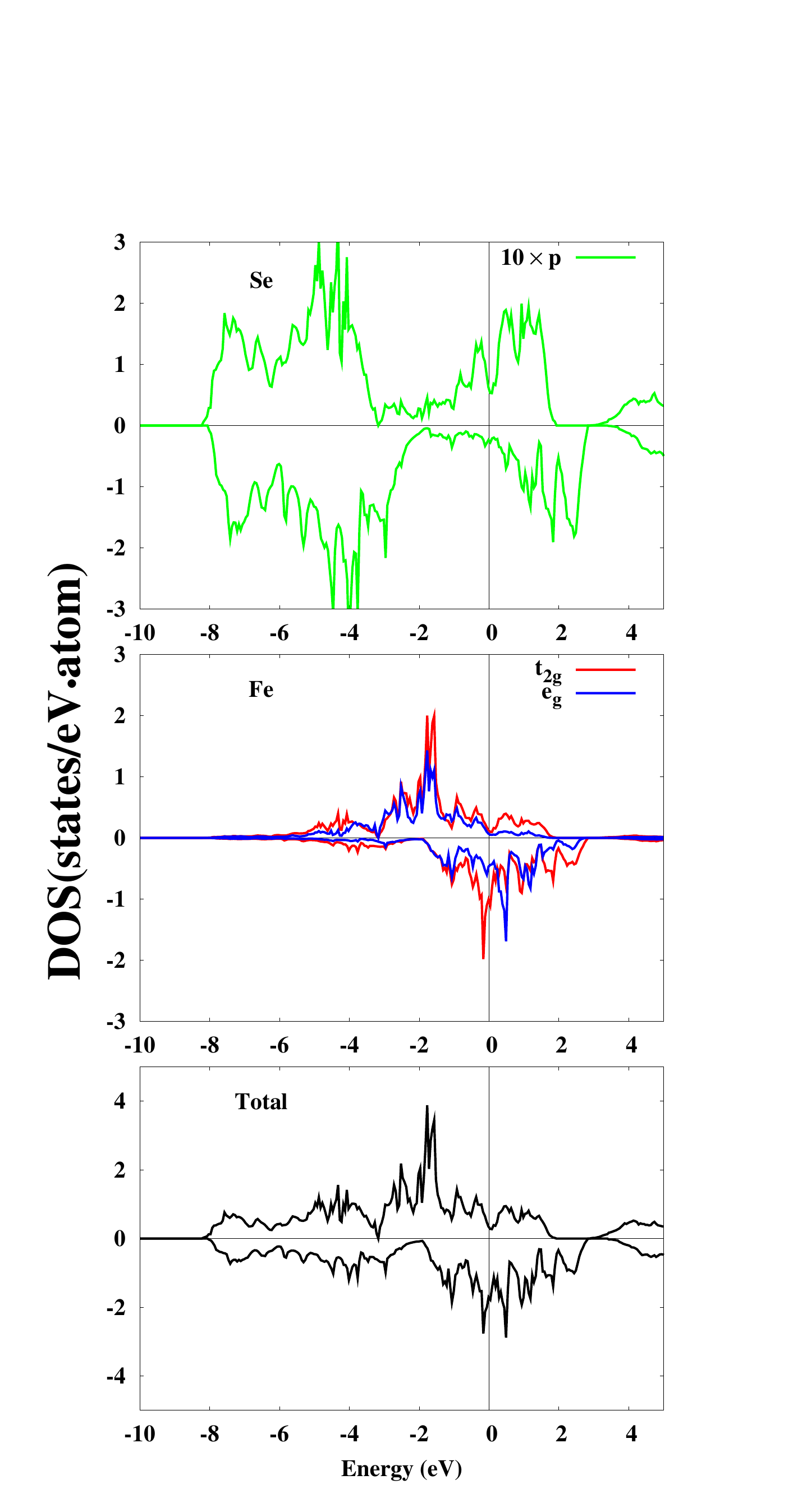}\\
\vfill
\textbf{Fig. 5(c)}
\end{center}

\newpage
\begin{center}
\vfill
\includegraphics[width=0.8\textwidth]{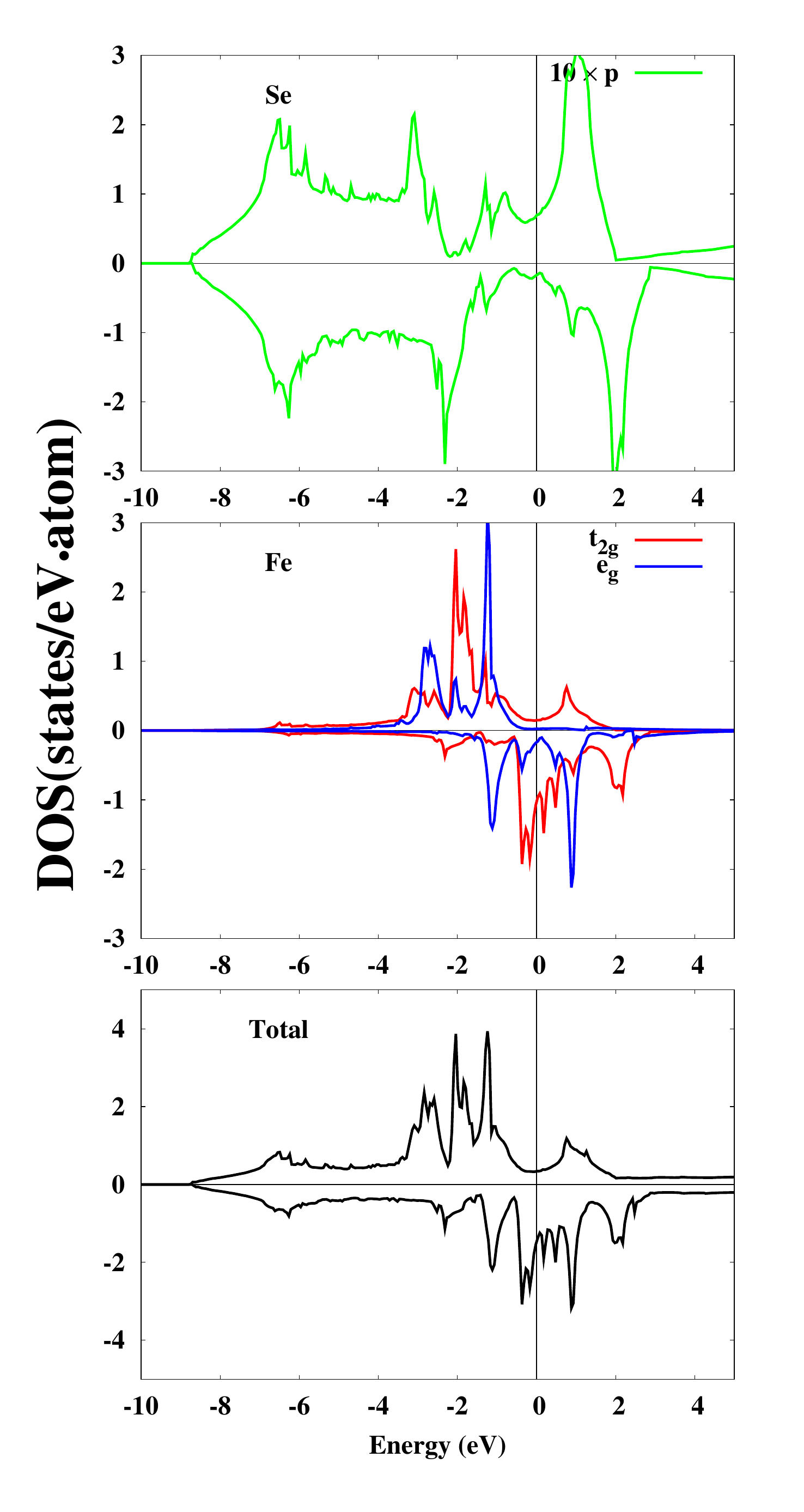}\\
\vfill
\textbf{Fig. 5(d)}
\end{center}

\newpage
\begin{center}
\vfill
\includegraphics[width=0.7\textwidth]{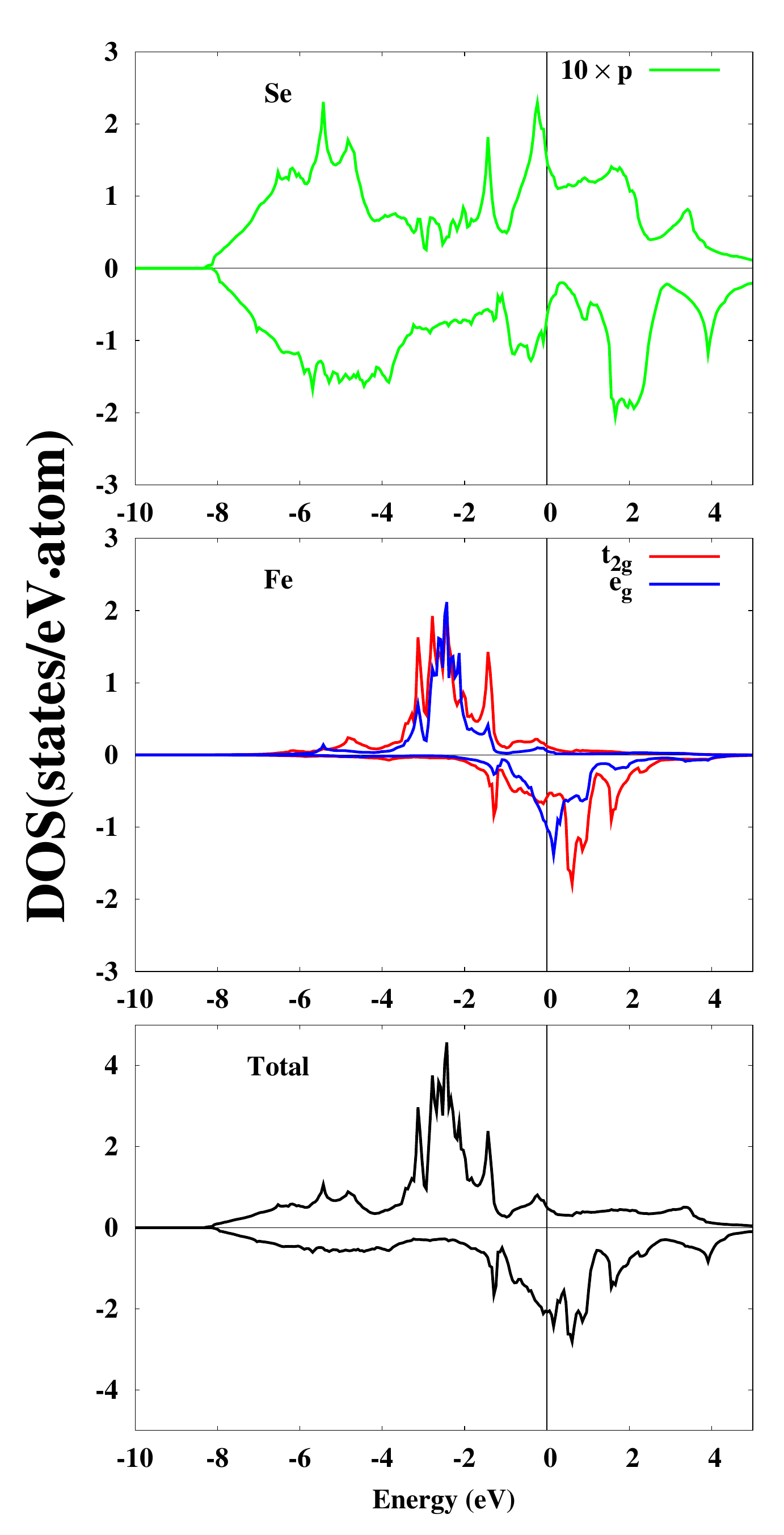}\\
\vfill
\textbf{Fig. 5(e)}
\end{center}

\newpage
\begin{center}
\vfill
\includegraphics[width=0.8\textwidth]{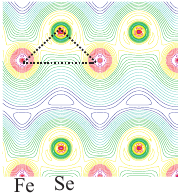}\\
\vfill
\textbf{Fig. 6(a)}
\end{center}

\newpage
\begin{center}
\vfill
\includegraphics[width=0.8\textwidth]{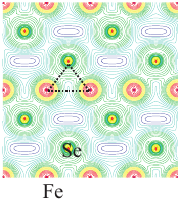}\\
\vfill
\textbf{Fig. 6(b)}
\end{center}

\newpage
\begin{center}
\vfill
\includegraphics[width=0.8\textwidth]{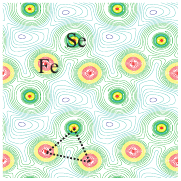}\\
\vfill
\textbf{Fig. 6(c)}
\end{center}

\newpage
\begin{center}
\vfill
\includegraphics[width=0.8\textwidth]{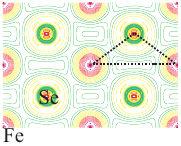}\\
\vfill
\textbf{Fig. 6(d)}
\end{center}

\newpage
\begin{center}
\vfill
\includegraphics[width=0.8\textwidth]{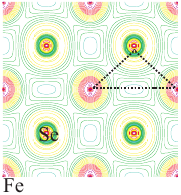}\\
\vfill
\textbf{Fig. 6(e)}
\end{center}

\end{document}